\documentclass[twocolumn]{aastex631}

\usepackage{amsmath}
\usepackage{multirow}
\usepackage{times}
\usepackage{xcolor}
\usepackage{url}
\usepackage{wrapfig}
\usepackage{graphicx}
\usepackage{bm}
\usepackage{comment}
\usepackage{longtable}

\received{April 28, 2023}
\revised{June 07, 2023}
\accepted{June 21, 2023}

\submitjournal{AJ}

\shorttitle{HST WFC3 Transmission Spectrum of LTT 9779 b}
\shortauthors{Edwards et al. 2023}

\graphicspath{{./}{Figures/}}

\begin{document}

\title{Characterising a World Within the Hot Neptune Desert: Transit Observations of LTT\,9779\,b with HST WFC3}

\correspondingauthor{Billy Edwards}
\email{b.edwards@sron.nl}

\author[0000-0002-5494-3237]{Billy Edwards}
\affil{SRON, Netherlands Institute for Space Research, Niels Bohrweg 4, NL-2333 CA, Leiden, The Netherlands}

\author[0000-0001-6516-4493]{Quentin Changeat}
\affil{European Space Agency (ESA), ESA Office, Space Telescope Science Institute (STScI), Baltimore MD 21218, USA}

\author[0000-0003-3840-1793]{Angelos Tsiaras}
\affil{INAF, Osservatorio Astrofisico di Arcetri, Largo E. Fermi 5, 50125 Firenze, Italy}

\author{Andrew Allan}
\affiliation{Leiden Observatory, Leiden University, PO Box 9513, 2300 RA Leiden, The Netherlands}

\author{Patrick Behr}
\affil{Laboratory for Atmospheric and Space Physics, University of Colorado Boulder, Boulder, CO 80303}

\author[0000-0001-8072-0590]{Simone R. Hagey}
\affil{The University of British Columbia, 6224 Agricultural Road Vancouver, BC V6T 1Z1, Canada}

\author[0000-0002-9338-8600]{Michael D.~Himes}
\affiliation{Planetary Sciences Group, Department of Physics, University of Central Florida, Orlando, FL, USA}
\affiliation{NASA Postdoctoral Program Fellow, NASA Goddard Space Flight Center, Greenbelt, MD, USA}

\author[0000-0001-9010-0539]{Sushuang Ma}
\affil{Department of Physics and Astronomy, University College London, Gower Street, London, WC1E 6BT, United Kingdom}

\author[0000-0002-3481-9052]{Keivan G.\ Stassun}
\affiliation{Department of Physics and Astronomy, Vanderbilt University, Nashville, TN 37235, USA}

\author{Luis Thomas}
\affil{Universit\"ats-Sternwarte M\"unchen, Scheinerstr. 1, D-81679 M\"unchen, Germany}
\affil{Max-Planck-Institut f\"ur extraterrestrische Physik, Giessenbachstrasse 1, D-85748 Garching, Germany}
\affil{Department of Physics and Astronomy, University College London, Gower Street, London, WC1E 6BT, United Kingdom}

\author{Alexandra Thompson}
\affil{Department of Physics and Astronomy, University College London, Gower Street, London, WC1E 6BT, United Kingdom}

\author{Aaron Boley}
\affil{The University of British Columbia, 6224 Agricultural Road Vancouver, BC V6T 1Z1, Canada}

\author{Luke Booth}
\affil{Cardiff Hub for Astrophysics Research and Technology (CHART), School of Physics and Astronomy, Cardiff University, UK}

\author{Jeroen Bouwman} 
\affil{Max-Planck-Institut für Astronomie, Königstuhl 17, 69117 Heidelberg, Germany}

\author{Kevin France}
\affil{Laboratory for Atmospheric and Space Physics, University of Colorado Boulder, Boulder, CO 80303}

\author[0000-0001-6508-5736]{Nataliea Lowson}
\affil{Centre for Astrophysics, University of Southern Queensland, 499-565 West Street, Toowoomba, QLD 4350, Australia}

\author[0000-0002-7500-7173]{Annabella Meech}
\affil{Department of Physics, University of Oxford, Denys Wilkinson Building, Oxford, OX1 3RH}

\author[0000-0001-5610-5328]{Caprice L. Phillips}
\affil{Department of Astronomy, The Ohio State University, 100 W 18th Ave, Columbus, OH 43210 USA}

\author[0000-0001-5371-2675]{Aline A.~Vidotto}
\affiliation{Leiden Observatory, Leiden University, PO Box 9513, 2300 RA Leiden, The Netherlands}

\author[0000-0002-9616-1524]{Kai Hou Yip}
\affil{Department of Physics and Astronomy, University College London, Gower Street, London, WC1E 6BT, United Kingdom}

\author{Michelle Bieger}
\affil{College of Engineering, Mathematics and Physical Sciences, University of Exeter, North Park Road, Exeter, United Kingdom}

\author[0000-0003-0854-3002]{Amélie Gressier}
\affil{Space Telescope Science Institute, 3700 San Martin Drive, Baltimore, MD 21218, USA}

\author{Estelle Janin}
\affil{Arizona State University, PO Box 871404, Tempe, 85287-1404, Arizona, USA}

\author[0000-0001-7359-3300]{Ing-Guey Jiang}
\affiliation{Department of Physics and Institute of Astronomy, National Tsing Hua University, Hsinchu 30013, Taiwan}

\author[0000-0001-6026-9202]{Pietro Leonardi}
\affil{Dipartimento di Fisica, Università di Trento, Via Sommarive 14, 38123 Povo, Italy}
\affil{Dipartimento di Fisica e Astronomia, Universit\`a degli Studi di Padova, Vicolo dell'Osservatorio 3, 35122 Padova, Italy}

\author{Subhajit Sarkar}
\affil{Cardiff Hub for Astrophysics Research and Technology (CHART), School of Physics and Astronomy, Cardiff University, UK}

\author[0000-0002-9372-5056]{Nour Skaf}
\affil{LESIA, Observatoire de Paris, Univ. Paris Cité, Univ.~PSL, CNRS, Sorbonne Univ., 5 pl. Jules Janssen, 92195 Meudon, France}
\affil{National Astronomical Observatory of Japan, Subaru Telescope, 650 North A'oh\=ok\=u Place, Hilo, HI 96720, U.S.A.}
\affil{Department of Physics and Astronomy, University College London, Gower Street, London, WC1E 6BT, United Kingdom}

\author{Jake Taylor}
\affil{Department of Physics (Atmospheric, Oceanic and Planetary Physics), University of Oxford, Parks Rd, Oxford OX1 3PU, UK}
\affil{Institut Trottier de Recherche sur les Exoplanètes and Département de Physique, Université de Montréal, 1375 Avenue Thérèse-Lavoie-Roux,\\ Montréal, QC, H2V 0B3, Canada}

\author[0000-0002-6926-2872]{Ming Yang}
\affiliation{College of Surveying and Geo-Informatics, Tongji University, Shanghai 200092, China}

\author{Derek Ward-Thompson}
\affil{Jeremiah Horrocks Institute, University of Central Lancashire, Preston PR1 2HE, UK}

\begin{abstract}

We present an atmospheric analysis of LTT 9779 b, a rare planet situated in the hot Neptune desert, that has been observed with HST WFC3 G102 and G141. The combined transmission spectrum, which covers 0.8 - 1.6 $\mu$m, shows a gradual increase in transit depth with wavelength. Our preferred atmospheric model shows evidence for H$_{\rm 2}$O, CO$_{\rm 2}$ and FeH with a significance of 3.1 $\sigma$, 2.4 $\sigma$ and 2.1 $\sigma$, respectively. In an attempt to constrain the rate of atmospheric escape for this planet, we search for the 1.083 $\mu$m Helium line in the G102 data but find no evidence of excess absorption that would indicate an escaping atmosphere using this tracer. We refine the orbital ephemerides of LTT\,9779\,b using our HST data and observations from TESS, searching for evidence of orbital decay or apsidal precession, which is not found. The phase-curve observation of LTT\,9779\,b with JWST NIRISS should provide deeper insights into the atmosphere of this planet and the expected atmospheric escape might be detected with further observations concentrated on other tracers such as Lyman $\alpha$. 

\end{abstract}


\section{Introduction} \label{sec:intro}

The exoplanet field has rapidly expanded, with thousands of planets currently known and thousands more anticipated in the coming decade, thanks to dedicated planet-hunting missions such as Kepler \citep{borucki} and the Transiting Exoplanet Survey Satellite \citep[TESS, ][]{ricker}. The vast number of detected worlds enables characterisation of a diverse exoplanet sample to be further characterised, with low resolution space based spectroscopy detecting broadband spectral features of molecular species in exoplanet atmospheres \citep[e.g.,][]{tinetti_water,swain2008methane}. 

For over a decade, the Hubble Space Telescope (HST) has been utilised to conduct observations of the atmospheres of exoplanets. With the spatial scanning technique allowing for higher signal-to-noise ratio observations \citep{mccullough_wfc3_scan}, the characterisation of exoplanetary atmospheres via the transit \citep[e.g.,][]{Kreidberg_GJ1214b_clouds,luque_w74,murgas_offset,skaf_ares,wilson_w103,yan_h12,guilluy_aresIV,yip_w96,mcgruder_w96,saba_w17}, eclipse \citep[e.g.,][]{line_hd209,edwards_ares,aresIII,changeat_edwards_k9,jacobs_k9}, and phase-curve \citep[e.g.,][]{stevenson_phase,Stevenson_2017,Kreidberg_w103,arcangeli_w18_phase,changeat_w43,robbins_w43,changeat_2021_w103} techniques has rapidly expanded. Indeed, enough planets have now been studied to facilitate population-style studies \citep[e.g.,][]{sing,tsiaras_30planets,fisher, cubillos_pop,mansfield_metric,roudier_diseq,emission_pop,transmission_pop}.

\begin{figure}
    \centering
    \includegraphics[width=\columnwidth]{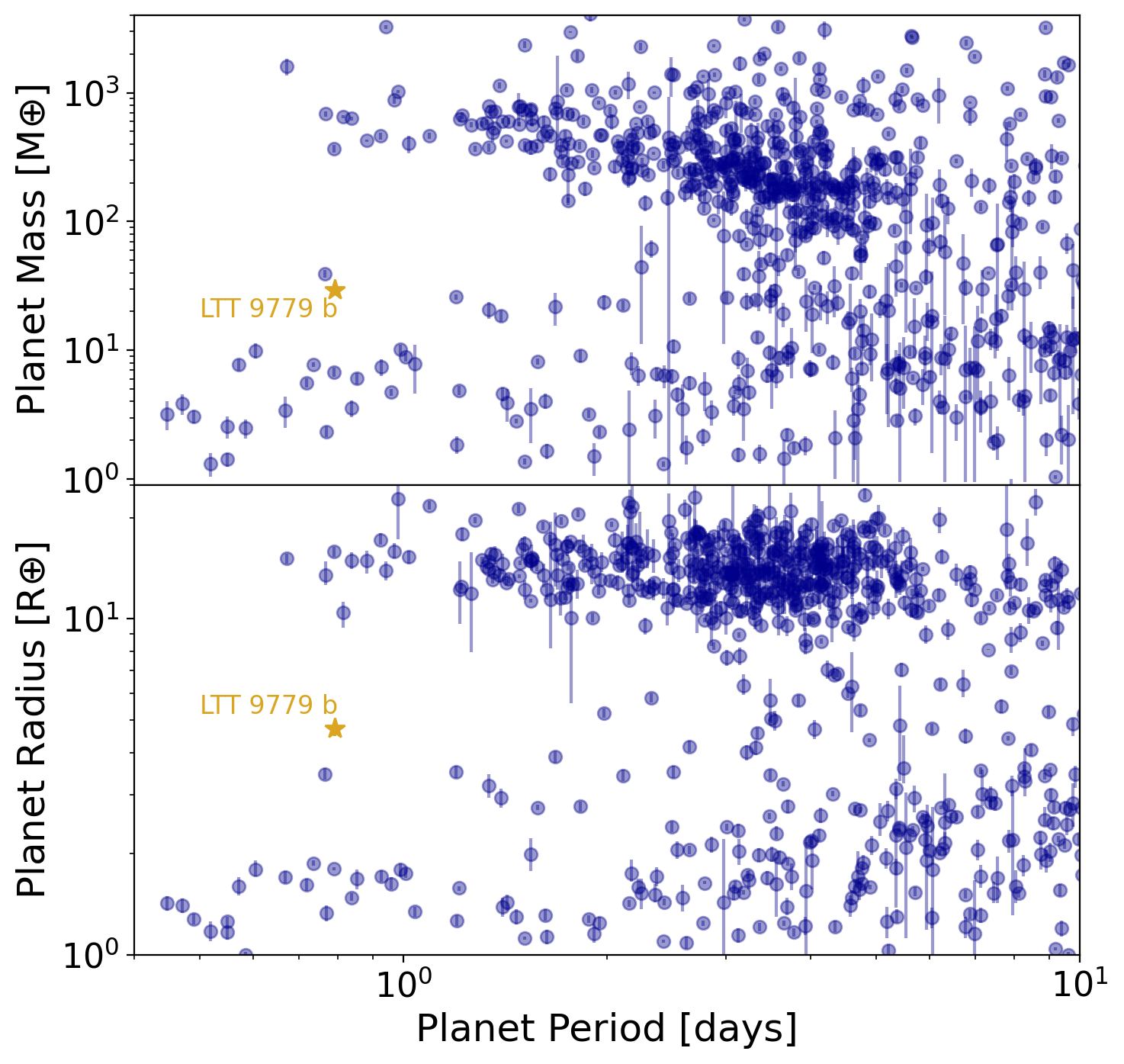}
    \caption{LTT\,9779\,b in context with the known population of transiting planets. In the period-mass and period-radius space, LTT\,9779\,b occupies a region which is sparsely populated. The majority of other planets on such a short orbital period are either stripped rocky cores or Jupiter-sized worlds.}
    \label{fig:context}
\end{figure}

Numerous short-period planets have been discovered and these usually fall into one of two populations: large, gaseous worlds (i.e. hot-Jupiters) or stripped rocky cores which are often referred to as ultra-short-period (USP) planets. Between these two populations, there is a dearth of planets and this area of the parameter space has been coined the hot Neptune desert \citep[e.g.,][]{szabo_nep_des,beauge_nep_des,mazeh_nep_des}. The lack of planets found in this region cannot be explained by observational biases due to the large number of Neptunian-sized worlds found on longer orbits. Instead, the under-population of this regime has been hypothesised to be caused by photoevaporation whereby planets that spend significant time in these short orbits either have enough mass to remain a Jupiter-sized world or else have their primordial atmosphere completely eroded \citep[e.g.,][]{lundkvist_evap,lopez_usp}.

LTT\,9779\,b \citep{jenkins_ltt9779} is an ultra-hot Neptune discovered using data from the TESS. With a period of less than a day and a radius of 4.7$\pm$0.23 R$_\oplus$, LTT\,9779\,b lies within the hot Neptune desert, as shown in Figure \ref{fig:context}. Radial velocity measurements have constrained the mass to 29$^{+0.78}_{-0.81}$ M$_\oplus$, yielding a density of 1.536 $\pm$0.123 g/cm$^{3}$ which is suggestive of a gas-rich world. Given the rarity of worlds residing within the hot Neptune desert, LTT\,9779\,b quickly became a target for atmospheric studies. Photometric eclipse observations of LTT\,9779\,b with Spitzer revealed a spectrum which is best-fitted by a non-inverted atmosphere and evidence was found for the presence of CO \citep{dragomir_ltt}. Phase curve observations with TESS and Spitzer have also revealed a large ($\sim$1100 K) day-night temperature contrast and have suggested a super-solar atmospheric metallicity \citep{crossfield_ltt}. However, the transit observations from these phase curves were not precise enough to constrain molecular composition of the atmosphere in transmission. As part of a homogeneous population study of 70 gaseous exoplanets, \citet{transmission_pop} presented the HST WFC3 G141 transmission spectrum of LTT\,9779\,b. They found a spectrum with few obvious features except a slope of increasing transit depth with wavelength across the 1.1 - 1.6~$\mu$m spectral range, which was best explained by a model with no molecular absorption, with the slope being instead best fit by Collision Induced Absorption (CIA) and Rayleigh scattering. Retrievals which included optical absorbers (TiO, VO, FeH, H-) suggested the presence of TiO, but this was found to be strongly dependent upon the bluest spectral point and thus it was noted this could simply be noise.

In this paper, we present a new set of observations from the HST WFC3 G102, reduced with the same methodology as in \cite{transmission_pop}. The combined HST WFC3 G102 and G141 transmission spectrum of LTT\,9779\,b covers the wavelengths from 0.8$\mu$m to 1.6$\mu$m. Analysing these data, our free chemistry atmospheric retrievals prefer solutions with high abundances of H$_2$O and CO$_2$, with a detection significance of 3.11 $\sigma$ and 2.35 $\sigma$, respectively, as well as the preferring the presence of FeH to 2.12 $\sigma$. The chemical equilibrium retrievals we conduct do not provide a preferable fit to the data than the free chemistry retrievals. Nevertheless, the best-fit solution suggests a carbon-to-oxygen ratio of C/O = 0.56$^{+0.41}_{-0.34}$ and a very low atmospheric metallicity of log(Z) = -2.74$^{+0.69}_{-0.56}$. Furthermore, we search for the 1.083 $\mu$m Helium line, but find no evidence for it within the HST data. Finally, we use the mid-times from the HST data, along with literature mid-times, to refine the period of LTT\,9779\,b and search for evidence of orbital decay or precession, finding no strong evidence that the orbital period is changing over the current observational baseline.

\section{Methodology}

HST WFC3 G102 and G141 observations of LTT\,9779\,b's transit were taken as part of proposal GO-16457 \citep[PI: Billy Edwards; ][]{edwards_prop}. These data were taken using the SPARS10 sequence and the GRISM256 aperture. The total exposure time was 103.13 seconds, with 16 up-the-ramp samples per exposure. Each visit consisted of four HST orbits, with the third occurring in-transit. For wavelength calibration, a direct image with the F132N filter was taken at the start of the first orbit of each observation sequence. The spatial scanning technique \citep{mccullough_wfc3_scan} was utilised to increase the duty cycle and, thus, improve the signal-to-noise ratio of the data. The scan rate employed was 0.18 "/s, yielding a total scan length of 19.7 ", which equated to around 150 pixels.

\subsection{Data Reduction and Light Curve Fitting}
\label{sec:data_red}

\begin{figure}
    \centering
    \includegraphics[width=\columnwidth]{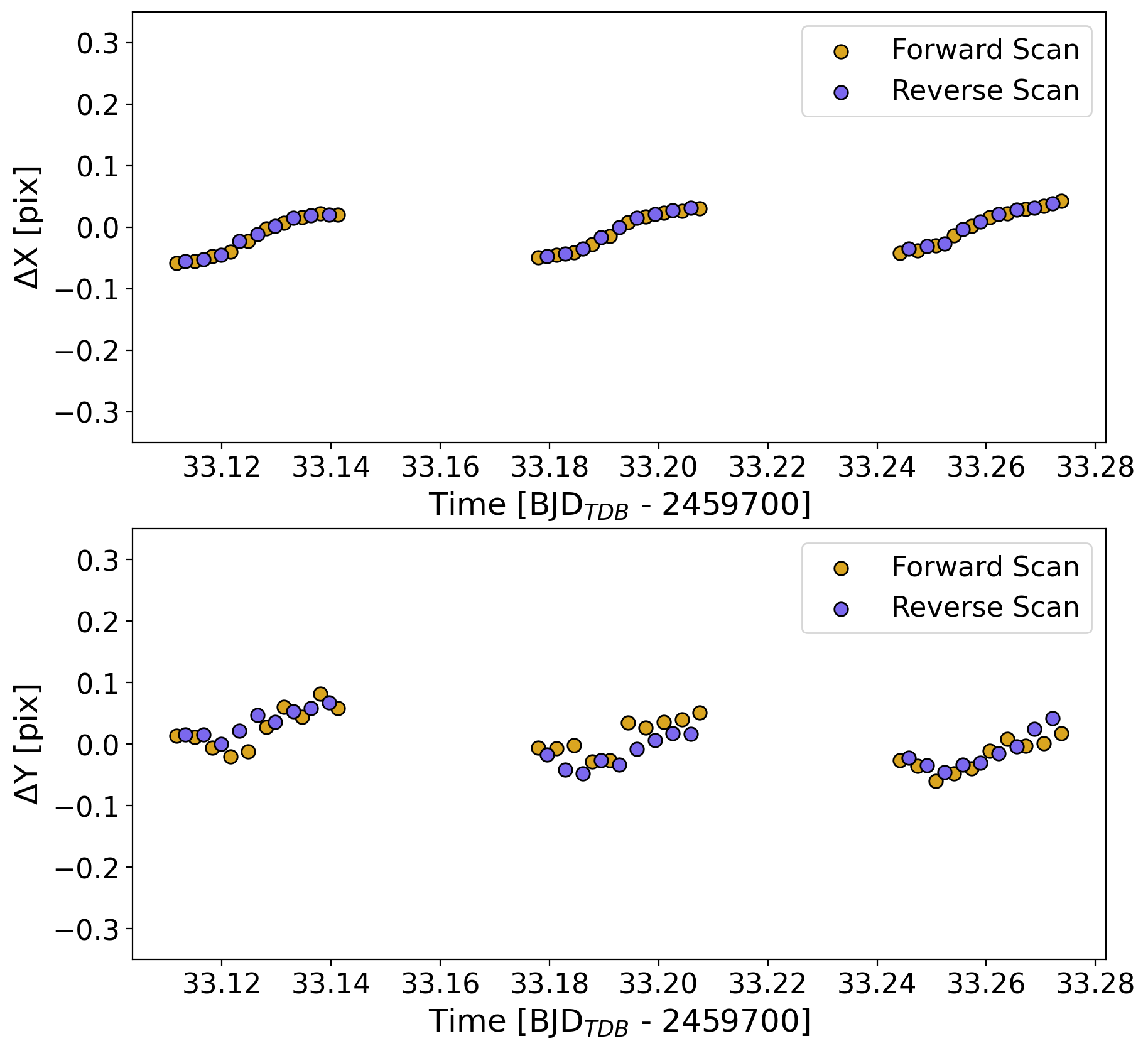}
    \includegraphics[width=\columnwidth]{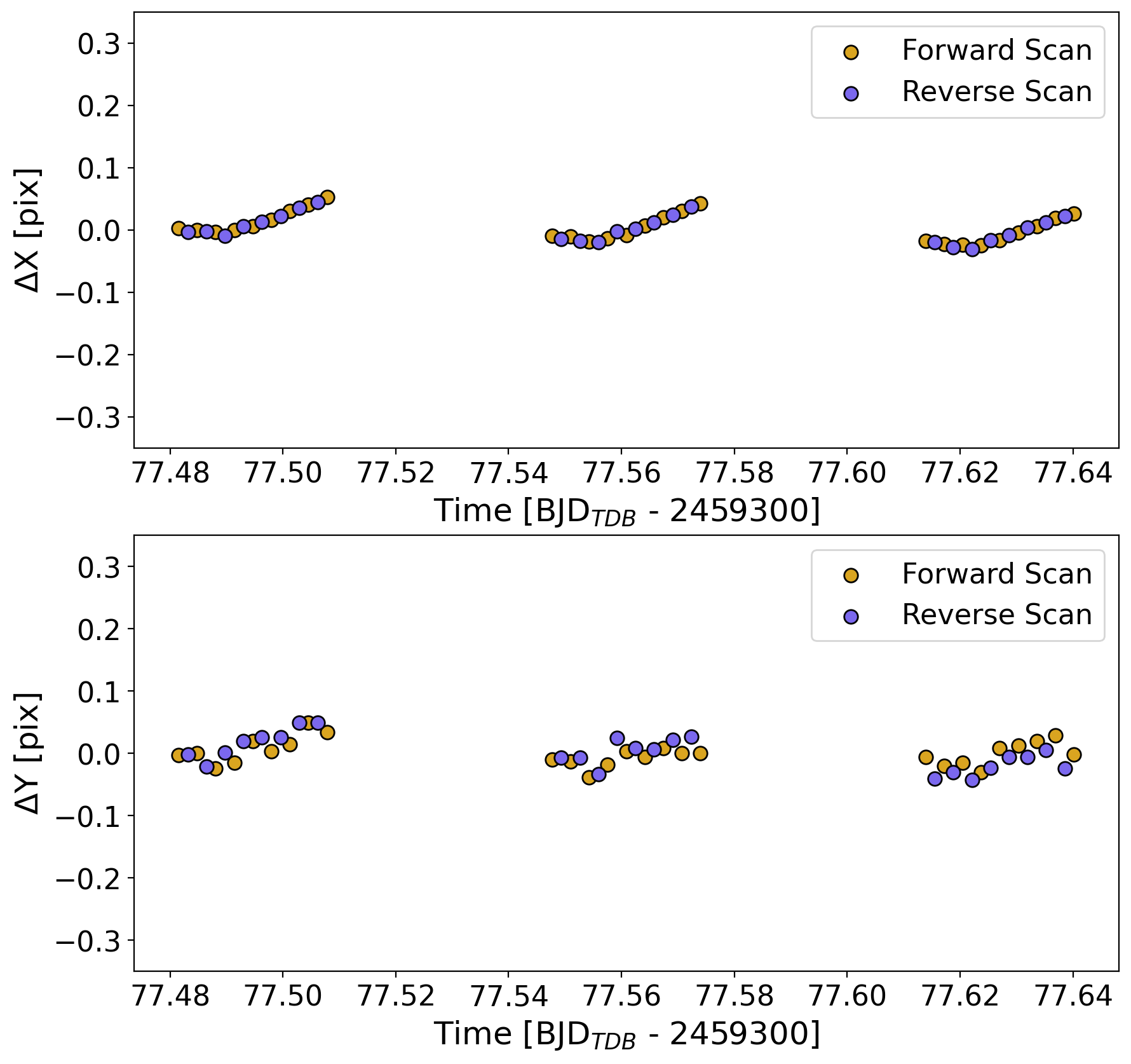}
    \caption{Shifts in the position of the spectrum on the detector in the spectral (x) and spatial (y) directions for the G102 (top) and G141 (bottom) visits. In both cases, the pointing across the visit was excellent.}
    \label{fig:shifts}
\end{figure}

\begin{figure*}
    \centering
    \includegraphics[width=\textwidth]{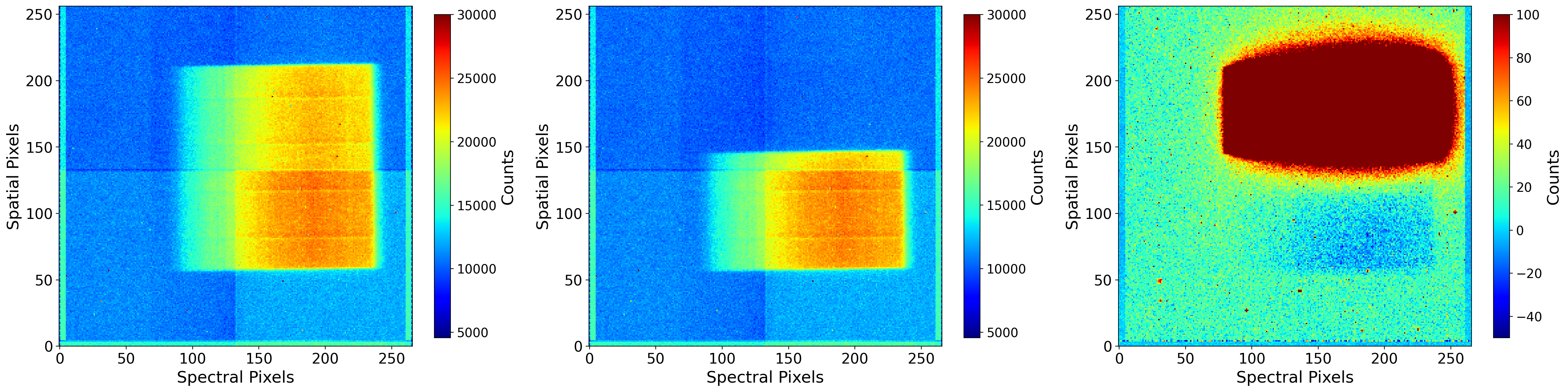}
    \includegraphics[width=\textwidth]{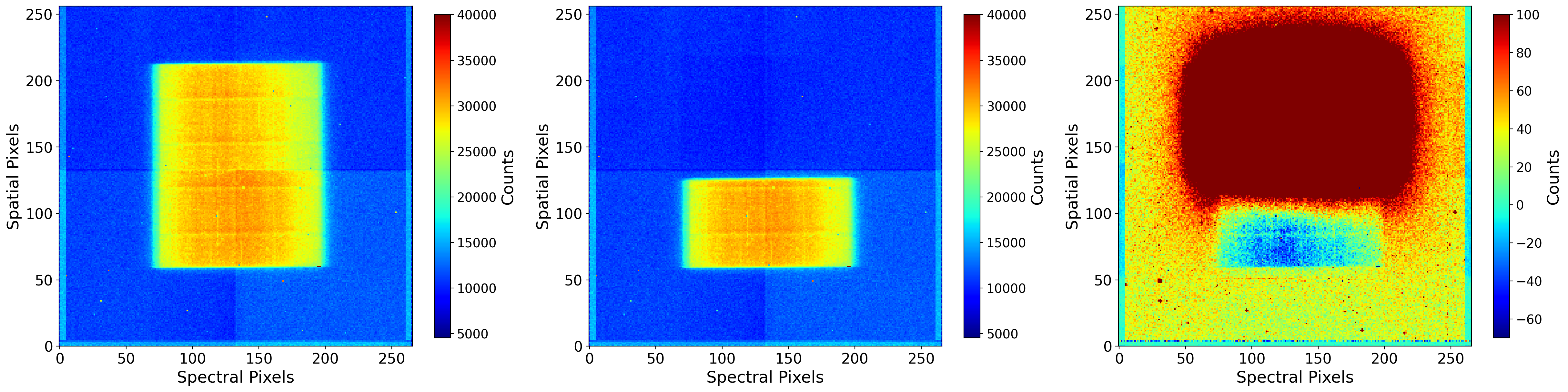}
    \caption{Raw images from the G102 (\textbf{top}) and G141 (\textbf{bottom}) observation sequences. In each case, the image illustrates the second forward scan of the second orbit, which is the first dataset used in this analysis. For each panel, the \textbf{left} image shows the final read while the \textbf{middle} image displays a non-destructive read. The \textbf{right} image shows the flux levels once the non-destructive read is subtracted. The regions of negative flux highlight the persistence effect for these datasets. The colour scale for the right images differ from the others in this Figure to highlight this negative region.}
    \label{fig:persistance}
\end{figure*}

We carry out the analysis of the transit data using \texttt{Iraclis}, our highly-specialised software for processing WFC3 spatially scanned spectroscopic images \citep{tsiaras_hd209,tsiaras_55cnce,tsiaras_30planets} which has been used in a number of studies \citep[e.g.,][]{anisman_w117,changeat_k11,edwards_lhs,libby_gj1132,brande_toi674,garcia_t1h}. The reduction process includes the following steps: zero-read subtraction, reference-pixels correction, non-linearity correction, dark current subtraction, gain conversion, sky background subtraction, calibration, flat-field correction and bad-pixels/cosmic-rays correction. Then we extract the white (1.088-1.680 $\mu$m) and the spectral light curves from the reduced images, taking into account the geometric distortions caused by the tilted detector of the WFC3 infrared channel. The pointing performance of HST was excellent across both visits and Figure \ref{fig:shifts} shows the shift in the X and Y position of the spectrum.

In the extraction of the flux from the calibrated images, \texttt{Iraclis} offers two techniques. The first extracts the flux from the image as it is at the end of the scan. The second splits the data into the up-the-ramp reads, subtracting each one sequentially from the next read and extracting the flux from each resulting image. The splitting extraction is useful for ensuring that there are no background stars with overlapping contributions to the target star's spectrum. However, differences between the flux extracted with each method can also occur due to persistence. While the level of persistence is dependent upon the accumulated charge, it is also correlated with the time the charge has spent on the detector as well as the time since the last read \citep{anderson_persistence}. 

In the context of HST WFC3, this means the persistence is dependent upon the brightness of the host star, the scanning rate, and the readout scheme employed. We show this effect for the second forward scan of the second orbit (the first image used in the analysis) in Figure \ref{fig:persistance}. The differences are clear when comparing the G102 datasets to those taken with the G141 grism: the persistence effect is greater in the latter case, likely due to the higher flux levels as the scan rate and readout schemes were identical. As the extracted flux using the full scan and splitting methods can be different, and to minimise the impact of persistence, we use the splitting extraction in this work.

\begin{figure*}
    \centering
    \includegraphics[width=0.475\textwidth]{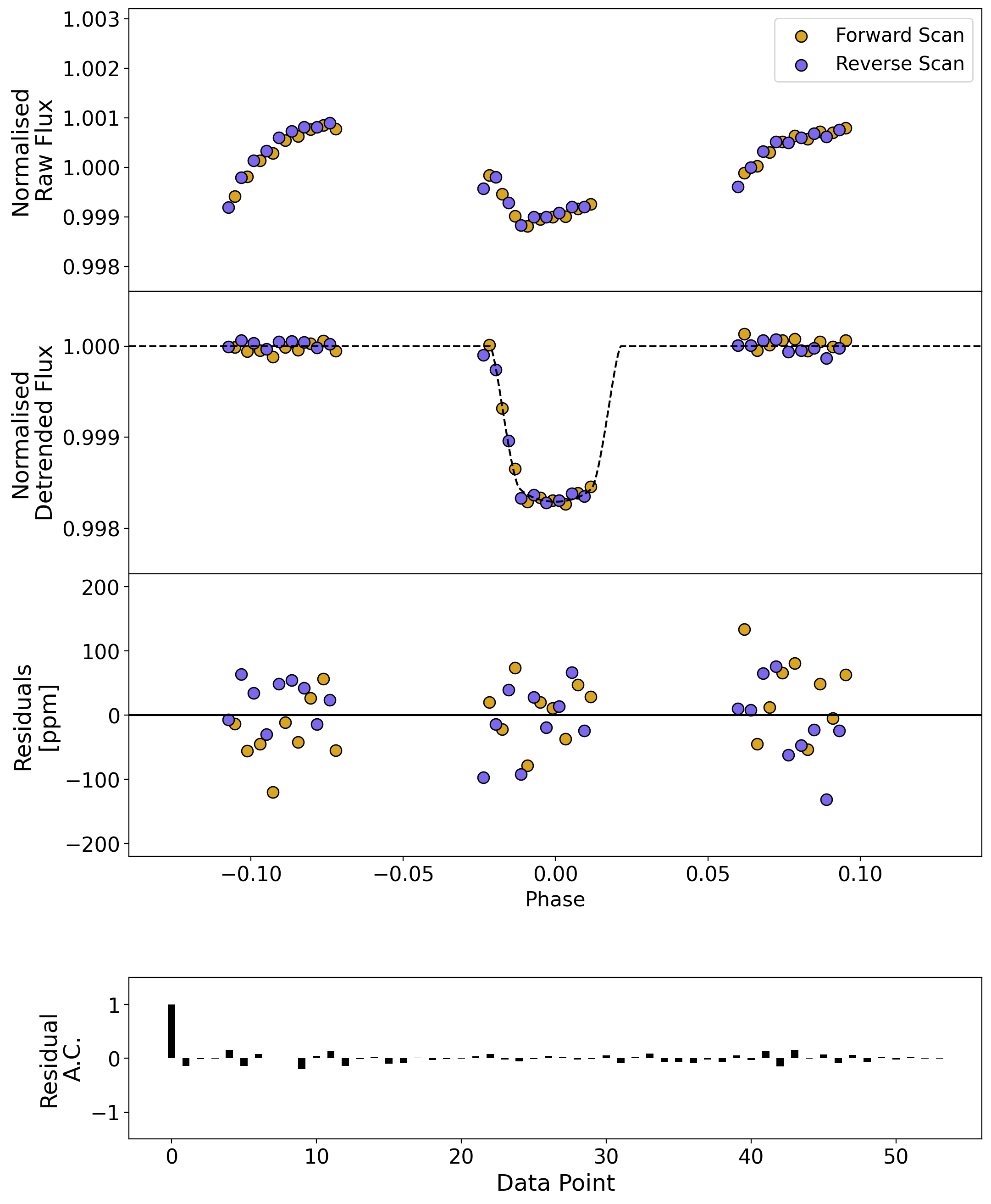}
    \includegraphics[width=0.475\textwidth]{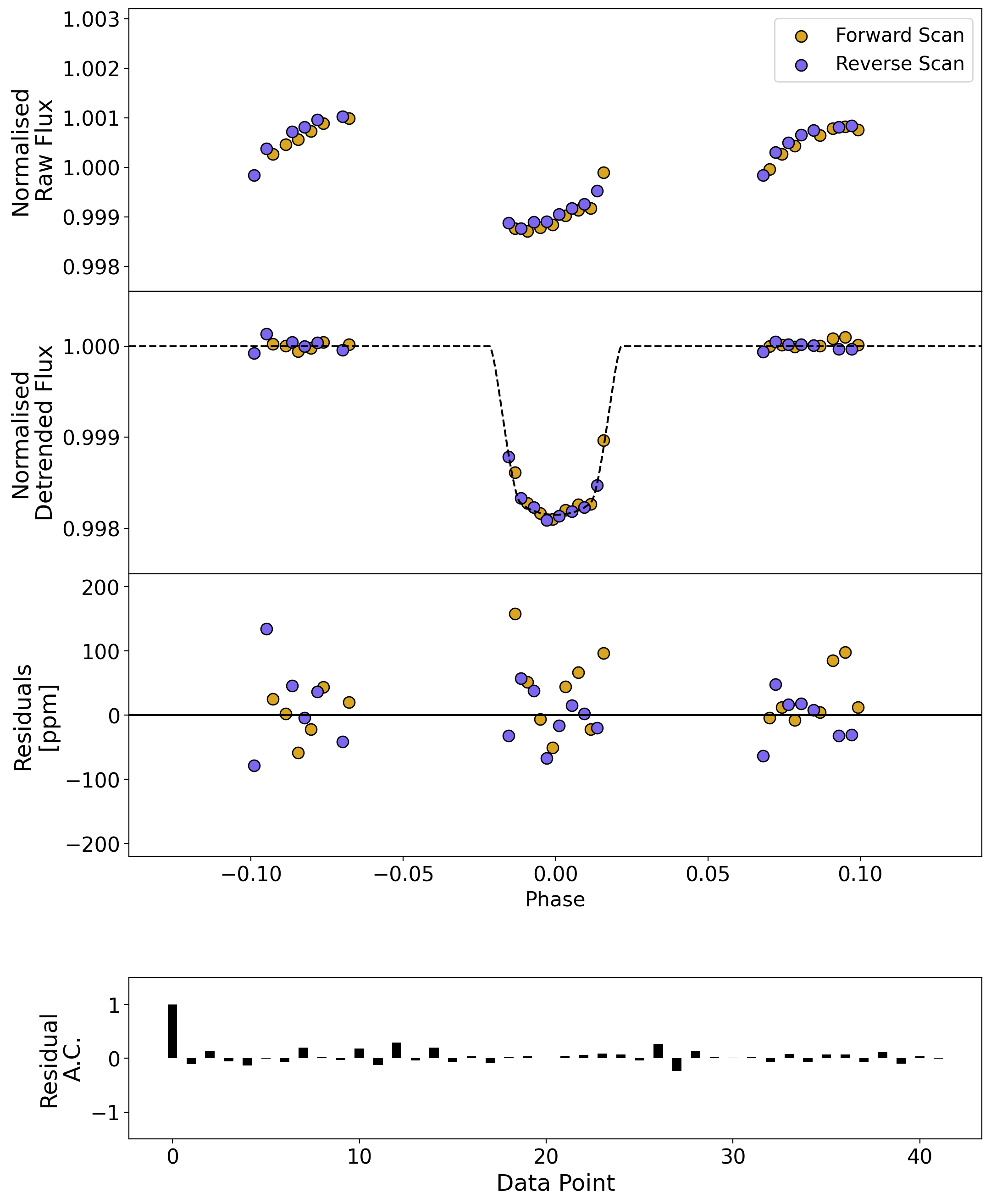}
    \includegraphics[width=0.475\textwidth]{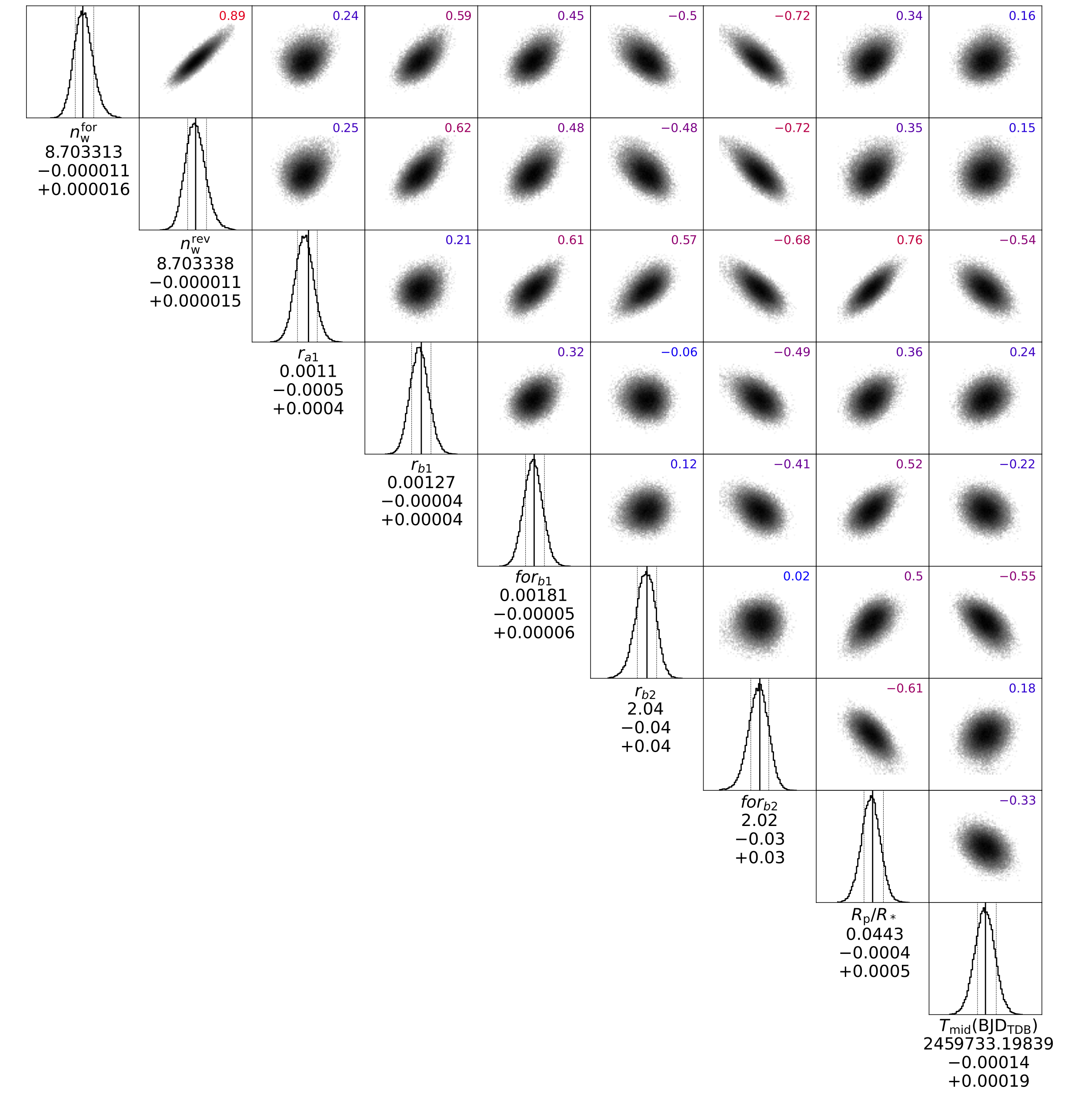}
    \includegraphics[width=0.475\textwidth]{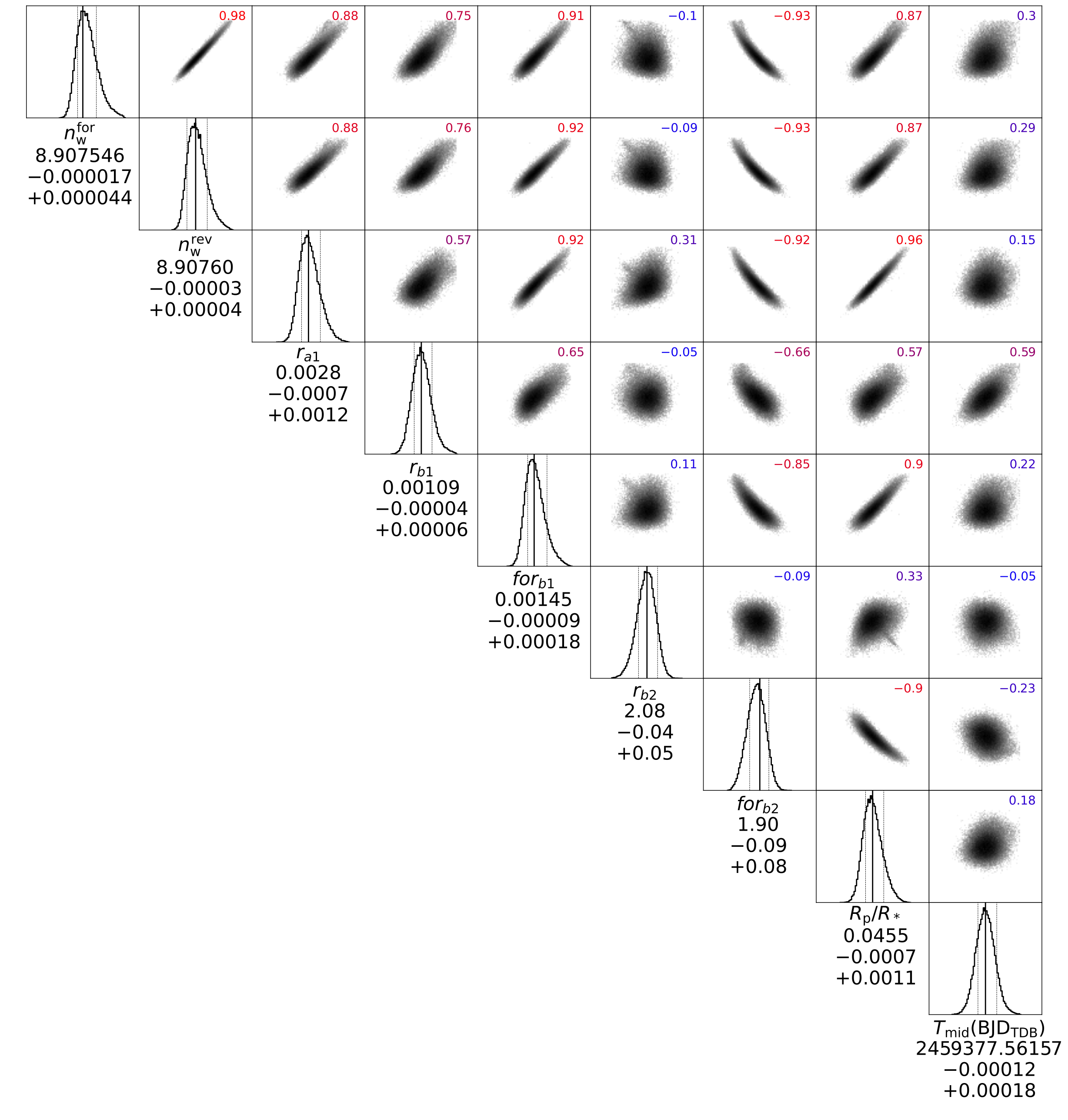}
    \caption{White light curve fits for the HST WFC3 G102 (\textbf{left}) and G141 (\textbf{right}) data of LTT\,9779\,b. In each case, the top plot shows the raw light curve (\textbf{top panel}), corrected light curve and best-fit model (\textbf{second panel}), the residuals having removed the best-fit model (\textbf{third panel}), and the auto-correlation function (\textbf{bottom panel}). The lower graphic is the corner plot for the fit. As with the majority of HST datasets, there is a correlation between the white light curve depth and the parameters for the systematics (Equation 1), thereby indicating that achieving absolute transit depths is difficult with this data. For the data studied here, the G141 data shows the highest correlation. Nevertheless, no major offset in the transit depth was found between the two visits.}
    \label{fig:white_lc}
\end{figure*}

We fit the light curves using our transit model package \texttt{PyLightcurve} \citep{tsiaras_plc} with the transit parameters from Table \ref{tab:para}. The limb-darkening coefficients are calculated using \texttt{ExoTETHyS} \citep{morello_exotethys} and based on the \texttt{PHOENIX} 2018 models from \citet{phoenix}. The stellar parameters used are also given in Table \ref{tab:para}.

During our fitting of the white light curve, the planet-to-star radius ratio (R$_{\rm p}$/R$_{\rm s}$) and the mid-transit time (T$_0$) were the only free parameters, along with the parameters used to model the systematics \citep{tsiaras_hd209}. It is common for WFC3 exoplanet observations to be affected by two kinds of time-dependent systematics: the long-term and short-term `ramps'. The first affects each HST visit and has generally modelled as a linear behaviour, while the second affects each HST orbit is modelled as having an exponential behaviour \citep[see e.g.,][]{Kreidberg_GJ1214b_clouds,tsiaras_hd209}. The parametric model we use for the white light curve systematics (R$_{\rm w}$) takes the form:

\begin{equation}
    R_w(t) = n^{scan}_w(1-r_a(t - T_0))(1-r_{b1}e^{-r_{b2}(t-t_o)}),
\end{equation}

\noindent where t is time, n$^{\rm scan}_{\rm w}$ is a normalisation factor, T$_0$ is the mid-transit time, t$_{\rm o}$ is the time when each HST orbit starts, r$_{\rm a}$ is the slope of a linear systematic trend along each HST visit and (r$_{\rm b1},r_{\rm b2}$) are the coefficients of an exponential systematic trend along each HST orbit. The normalisation factor we use (n$^{\rm scan}_{\rm w}$) is changed to n$^{\rm for}_{\rm w}$ for upward scanning directions (forward scanning) and to n$^{\rm rev}_{\rm w}$ for downward scanning directions (reverse scanning). The reason for using separate normalisation factors is the slightly different effective exposure time due to the known upstream/downstream effect \citep{mccullough_wfc3_scan}. 

We fit the white light curves using the formulae above, considering the uncertainties per pixel as propagated through the data reduction process. However, it is common in HST/WFC3 data to have additional scatter that cannot be captured by the ramp model. For this reason, we apply a scaling factor to each of the individual data point uncertainties to match their median to the standard deviation of the residuals and repeated the fitting \citep{tsiaras_30planets}. The resulting fits for both the G102 and G141 observations are shown in Figure \ref{fig:white_lc}.

\begin{figure*}
    \centering
    \includegraphics[width=0.47\textwidth]{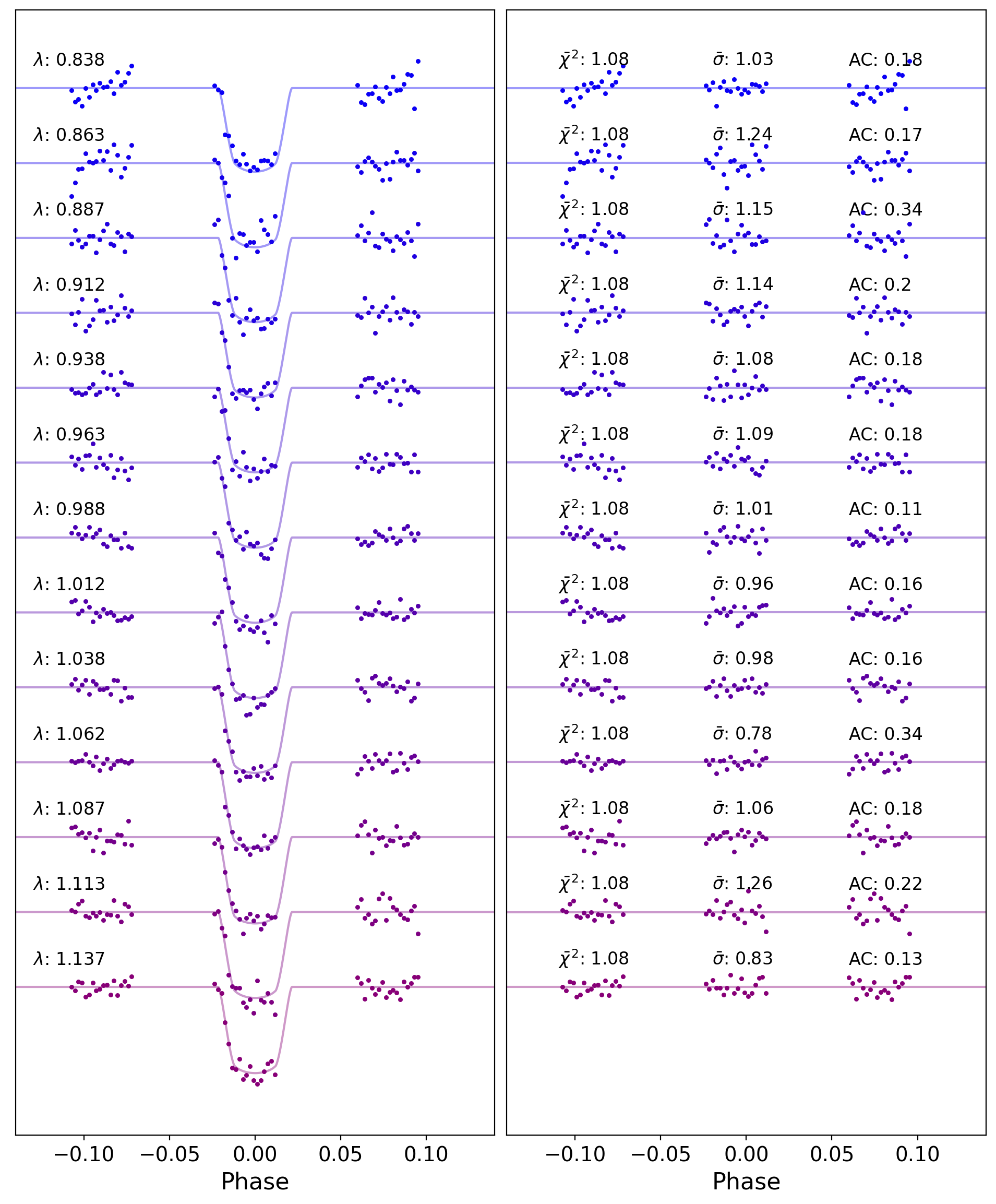}
    \includegraphics[width=0.47\textwidth]{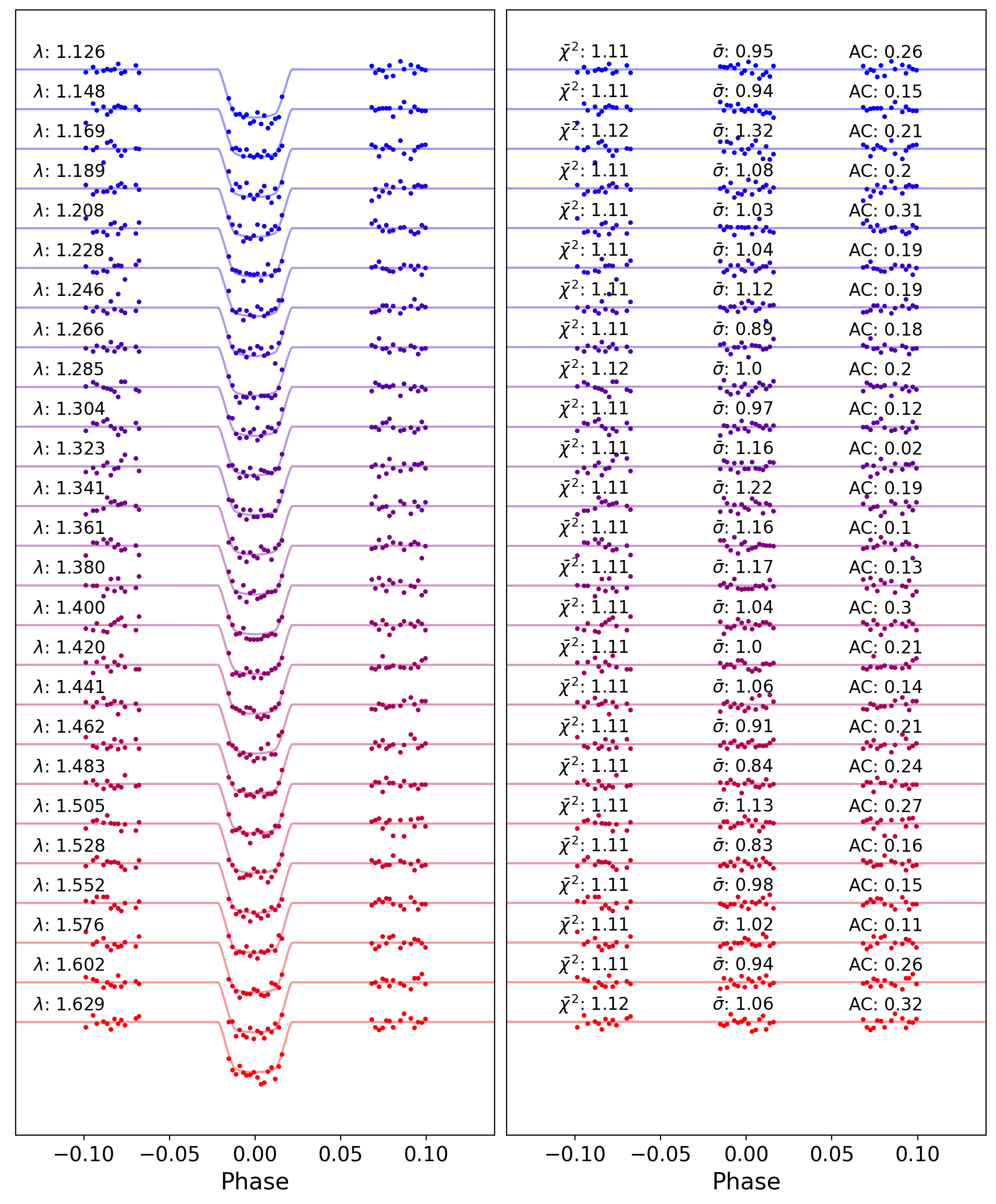}
    \caption{Spectral light curves fitted with \texttt{Iraclis} for the G102 (\textbf{left}) and G141 (\textbf{right}) observations of LTT\,9779\,b where, for clarity, an offset has been applied. Each plots shows the detrended spectral light curves with best-fit model plotted (\textbf{left}) as well as the residuals from the fitting (\textbf{right}). The values for the reduced Chi-squared ($\bar{\chi}^2$), the standard deviation of the residuals with respect to the photon noise ($\bar{\sigma}$), and the auto-correlation (AC) function are also shown.}
    \label{fig:spec_lc}
\end{figure*}

Next, we extract spectral light curves using the Iraclis ``default\textunderscore high" binning for both grisms. We fit these spectral light curves with a transit model. In this case, the free parameters are the planet-to-star radius ratio and the parameters for the systematics model. The mid-time was fixed to the best-fit value from the white light curve fit while other orbital parameters were again fixed to those in Table 2. The systematics model, ($R_\lambda$), includes the white light curve and a wavelength-dependent, visit-long slope \citep{tsiaras_hd209}, taking the form: 
\begin{equation}
    R_\lambda(t) = n^{scan}_\lambda(1-\chi_\lambda(t-T_0))\frac{LC_w}{M_w},
\end{equation}{}
\noindent where $\chi_\lambda$ is the slope of a wavelength-dependent linear systematic trend along each HST visit, LC$_{\rm}w$ is the white light curve and $M_w$ is the best-fit model for the white light curve. Again, the normalisation factor we use (n$^{\rm scan}_\lambda$) is changed to n$^{\rm for}_\lambda$ or n$^{\rm rev}_\lambda$ for upward or downward scanning directions respectively. Also, in the same way as for the white light curves, we perform an initial fit using the pipeline uncertainties and then refit while scaling-up these uncertainties, such that their median matches the standard deviation of the residuals. The spectral light curves fits, for both visits, are given in Figure \ref{fig:spec_lc}.

\subsection{Atmospheric Retrievals}

We explore the nature of LTT\,9779\,b's lower atmosphere using Bayesian retrievals. These atmospheric retrievals are performed on the extracted transmission spectrum using the publicly available retrieval suite \texttt{TauREx 3.1} \citep{al-refaie_taurex3,taurex3_chem}\footnote{\url{https://github.com/ucl-exoplanets/TauREx3_public}}. In our retrievals, we assume that LTT\,9779\,b possesses a primary atmosphere with a solar ratio of helium to hydrogen (He/H$_2$ = 0.17). The atmosphere of LTT\,9779\,b is simulated to range from 10$^{-4}$ to 10$^6$ Pa (10$^{-9}$ to 10 Bar) and sampled uniformly in log-space by 100 atmospheric layers. The retrieved radius is, therefore, the 10 Bar radius. We include Collision Induced Absorption (CIA) from H$_2$-H$_2$ \citep{abel_h2-h2, fletcher_h2-h2} and H$_2$-He \citep{abel_h2-he} as well as Rayleigh scattering for all molecules. We model clouds as a uniform opaque deck, fitting only the cloud-top pressure (i.e. grey clouds). We then perform two main types of retrievals, those which fit each molecular species independently (i.e. free chemistry) and those which assume molecules to be in chemical equilibrium. 

For the free chemistry retrievals, we include the molecular opacities from the ExoMol \citep{Tennyson_exomol}, HITRAN \citep{gordon} and HITEMP \citep{rothman} databases. Based on the expected chemical species of such a hot planet, we include the opacities of H$_2$O \citep{polyansky_h2o},  CO \citep{li_co_2015}, CO$_2$ \citep{rothman_hitremp_2010}, HCN \citep{Barber_2013_HCN},  TiO \citep{McKemmish_TiO_new}, VO \citep{mckemmish_vo}, FeH \citep{wende_FeH} and H$^-$ \citep{john_1988_h-}. To implement the last of these, we use the methodology described in \citet{edwards_ares}, fixing the neutral hydrogen volume mixing ratio and thus only fitting for e- volume mixing ratio. We note that \citet{himes_harrington_h-} showed that Equation 3 from \citet{john_1988_h-} does not replicate the table of coefficients in that same paper. \citet{himes_harrington_h-} recommend using Table 1 from \citet{john_1988_h-}, which is what is implemented here.

In each free chemistry case, all molecular abundances are allowed to vary from log(VMR) = -1 to log(VMR) = -15. Higher mixing are not expected in the hydrogen-dominated atmosphere of LTT\,9779\,b and, in any case, these would also necessitate accounting for self-broadening of the molecular lines \citep{anisman_lb2,anisman_lb}. We explore using both an isothermal temperature profile and an Npoint profile. For the former, we fit only the temperature. For the later, we fit the temperature at the ``surface'' (10 Bar) and the temperature at the ``top'' of the atmosphere (10$^{-9}$ Bar) as well as two intermediary points. For these intermediary points, we also fit the pressure at which these points occur. Therefore, the isothermal profile adds only a single free parameter to the retrieval while the NPoint adds six. For the free chemistry retrievals, we explore both chemical profiles which were constant with altitude and more complex, varying with altitude, two-layer profiles \citep{Changeat_2019}. 

Additionally, we conduct equilibrium chemistry retrievals using the code \texttt{GGchem} \citep{woitke_ggchem} via the \texttt{TauREx} plugin system \citep{taurex3_chem}. As with the free chemistry retrievals, we include Rayleigh scattering and CIA as well as simple grey clouds. For these retrievals, the free chemical parameters are the atmospheric metallicity and C/O ratio. For the equilibrium chemistry retrievals, the metallicity is allowed to vary from 1e-4 to 1e4 and the C/O ratio has bounds of 1e-3 and 2.

Recovering absolute transit depths is difficult, particularly given the strong systematics present within HST observations \citep[e.g., ][]{guo_hd97658,changeat_k11}. Therefore, one can induce offsets in the transit depth between datasets from different instruments\footnote{Offsets are also found between different observations with the same instrument, see e.g., \citet{transmission_pop}.} which, if left uncorrected, could bias the retrieved atmospheric composition. Hence, we also conduct retrievals where we allow the G141 dataset to be shifted relative to the G102 data, with this offset being a free parameter in the retrieval \citep[e.g., ][]{luque_w74,yip_w96}. We allow the G141 data to be shifted due to the strong correlations between the transit depth and systematics model (see Figure \ref{fig:white_lc}). To allow for this possibility to be fully explored, the bounds for this offset are set to be extremely broad: $\pm$1e-2 (10000 ppm).

Stellar activity, in the form of photospheric heterogeneities (i.e., spots and faculae), can contaminate the transmission spectrum of an exoplanet \citep[e.g., ][]{ballerini,mccullough_star,rackham,barclay_2021}. The contamination introduced is strongly chromatic with the strongest effects seen at shorter, optical wavelengths. As such, the possibility of stellar contamination is particularly important to consider for the G102 observations. Hence, we conduct retrievals using the \texttt{ASteRA} (Active Stellar Retrieval Algorithm) plugin for \texttt{TauREx}, outlined in \citet{thompson_ret_spots}, to explore potential contamination in the case of LTT\,9779\,b. In our setup, we allow for possible contamination effects due to both unocculted spots and unocculted faculae. \texttt{ASteRA} requires 4 additional fitting parameters which are the spot and faculae temperatures, T$_{\rm Spot}$ and T$_{\rm Fac}$ respectively, and their respective filling factors, F$_{\rm Spot}$ and F$_{\rm Fac}$. We again use intentionally nonrestrictive priors, with F$_{\rm Spot}$ and F$_{\rm Fac}$ bounds of 0 and 0.99, T$_{\rm Spot}$ bounds of 4000K and 5440K and T$_{\rm Fac}$ bounds of 5450K to 6000K.

Finally, we explore the parameter space using the nested sampling algorithm \texttt{Multinest} \citep{Feroz_multinest,buchner_multinest} with 1000 live points and an evidence tolerance of 0.5. We then utilise the Bayesian evidence derived by Multinest as a means of selecting the preferred atmospheric model and for judging the significance of molecular detections.

\subsection{Probing Atmospheric Mass Loss via the Helium Line at 1.083 $\mu$m}
\label{meth:he_line}

Close-in gaseous planets are expected to be undergoing atmospheric escape due to the high levels of stellar irradiation. Indeed, hydrodynamic escape has been proposed as the cause of the dearth of hot Neptunes, with planets either being completely stripped or having a high-enough starting mass to retain the majority of their primordial envelope \citep[e.g., ][]{lundkvist_evap,mazeh_nep_des}. 

Observational studies have been undertaken in an attempt to detect atmospheric escape and constrain the rate of mass loss. Many of these have focused on using the Ly$\alpha$ line, but despite these successes, attenuation by neutral hydrogen in the interstellar medium removes the majority of the Ly$\alpha$ line profile and so other tracers of mass-loss have also been proposed. One such alternative line is that of metastable Helium which occurs at 1.083$\mu$m \citep{oklopcic_he}. Ground-based observations have had notable success in detecting escape via this observable \citep[e.g.,][]{allart_h11,nortmann_w69,zhang_helium}, but the first detection of Helium came via observations with HST. The G102 grism of WFC3 was used to observed WASP-107\,b \citep{anderson_w107}, with an excess absorption being detected around 1.083 $\mu$m \citep{spake_w107}. These data suggested a mass-loss rate of $10^{10}$ - $3\times 10^{11}$ g s$^{-1}$. Similarly, HST WFC3 G102 data of HAT-P-11\,b found evidence for a mass-loss rate of 10$^{9}$ - 10$^{11}$ g s$^{-1}$ \citep{mansfield_h11} although high-resolution ground-based observations suggested a lower mass-loss rate \citep{allart_h11}. Observations of the Helium triplet have also been undertaken for planets on the edge of the hot Neptune desert, with the resulting mass-rates suggesting that the upper-edge (i.e., the higher mass planets) of the desert is impervious to atmospheric escape \citep{vissapragada_helium}.

As LTT\,9779\,b lies within the rarely populated hot Neptune desert it is expected that atmospheric loss should be ongoing, despite the old age of the system. Therefore, as well as using the HST WFC3 G102 data to constrain the atmospheric composition via a low-resolution spectrum, we also extract a higher-resolution spectrum to search for evidence of mass-loss via the Helium line. Using the excellent wavelength calibration offered by \texttt{Iraclis}, we extract overlapping bins with a bandwidth of 3 nm (30 \r{A}) and a separation of 0.1 \r{A} across the entire G102 range and perform fits to each of these light curves in the same way as discussed in Section \ref{sec:data_red}. We then use the transit depths recovered in these overlapping bins to search for excess absorption around 1.083$\mu$m.

Separately, we estimate the expected atmospheric escape rate using a photo-evaporation model (Allan et al.~2023, submm.). In our model, the high-energy stellar radiation ionises hydrogen and helium atoms, and the excess energy released after the ionisation heats the planet's upper atmosphere, which more easily evaporate. We construct the incoming high-energy stellar spectrum of LTT\,9779 using HST STIS NUV/Optical observations (G230L and G430L) that were taken as part of the MUSCLES Extension for Atmospheric Transmission Spectroscopy programme \citep[GO-16166; PI: Kevin France; ][]{france_prop}. As LTT\,9779 is a Sun-like star \citep{jenkins_ltt9779}, the FUV portion of the spectrum (1170 - 2200 \r{A}) is assumed to be the spectrum of the quiet Sun \citep{woods_solar_spec}, scaled to the HST STIS G230L spectrum of LTT\,9779; the adopted scaling factor was 9.0519$\times10^{-16}$. Meanwhile, the HST STIS G140M data is utilised to reconstruct the Ly$\alpha$ using the methodology described in \citet{youngblood_ly_a}. The Ly$\alpha$ flux is then used to compute the EUV  (100 - 1170 \r{A}) flux of LTT 9779 using the scaling relations from \citet{linsky_2014}. Finally, as X-ray observations with Chandra led to a non-detection, we also use a scaled solar X-ray spectrum (5 - 100 \r{A}).


Our atmospheric escape model requires as input the incident flux at four energy bins,  X-ray (0.517-1.24 nm), hard EUV (10-36nm), soft-EUV (36-92 nm) and FUV+NUV (91.2-320 nm). By integrating the spectrum above, the luminosities at each of these bins are  3.35$\times 10^{26}$, 3.02$\times 10^{27}$, 2.30$\times 10^{27}$, and 2.22$\times 10^{31}$ erg s$^{-1}$, respectively. Our model then solves for the hydrodynamics equations (conservation of mass, momentum and energy) and ionisation balance of hydrogen \citep{allan2019}. In our most recent model version (Allan et al, submm.), we also self-consistently compute the helium population (neutral helium in the singlet and triplet states, as well as ionised helium), and their corresponding energetics. The helium abundance is an input parameter of our model and we run two models, the first assuming an abundance by number of $2\%$ and a second at 10\%. 

\subsection{Updated Ephemeris and Search for Orbital Decay or Precession}

Maintaining exoplanet ephemerides is crucial for ensuring that atmospheric studies of exoplanets can be undertaken. Therefore, we utilise the mid-times from our HST light curve fits, as well as literature observations, to refine the period of LTT\,9779\,b. Data from TESS Sector 2 led to the discovery of LTT\,9779\,b and it has since re-observed the host star in Sector 29.\footnote{We note that LTT 9779 will also be studied in TESS Sector 69 (Aug-Sep 2023).} These TESS data were analysed by \citet{kokori_3} as part of the ExoClock project \citep{kokori,kokori_2}, which aims to refine the ephemerides of planets which will be studied by the Ariel mission \citep{tinetti_ariel,edwards_ariel,tinetti_ariel2}. In our work, we took the TESS mid-times from this previous study and incorporated new unpublished TESS eclipses. As with \citet{kokori_3}, we utilise the 2 minute cadence Pre-search Data Conditioning (PDC) light curves \citep{smith_pdc,stumpe_pdc1,stumpe_pdc2}. Due to the poor signal-to-noise ratio of the TESS eclipses, we fit them in groups of four eclipses to achieve eclipse mid-time uncertainties that were of a similar precision to the transit measurements. Furthermore, before the end of operations, Spitzer observed eclipses and phase-curves of LTT\,9779\,b \citep{crossfield_ltt,dragomir_ltt}. We combine the mid-times from these studies with our data to extend the number of epochs during which LTT\,9779\,b has been observed. A full list of the mid-times utilised here are given in the Appendix in Tables \ref{tab:tr_mt} \& \ref{tab:ec_mt}.

To further constrain the orbital period, we utilise the radial velocity data taken of LTT\,9779 by CORALIE and HARPS which were presented in the detection paper of the planet \citep{jenkins_ltt9779}. To model the radial velocity profile, we use \texttt{RadVel} \citep{fulton_radvel}. The radial velocity data and the transit mid-times are fitted under one likelihood using \texttt{Multinest} \citep{Feroz_multinest,buchner_multinest} to explore the parameter space. For the linear ephemeris model, we assume a circular orbit (e = 0) following the conclusions in \citet{jenkins_ltt9779}.

Given the short period of LTT\,9779\,b, one might expect that the planet's orbit should be shrinking gradually over time due to a transfer of angular momentum from the planet to the host star, eventually leading to planetary engulfment. It may occur when the orbital period of a planet is shorter than the rotational period of its host star, as is the case for LTT\,9779\,b \citep{jenkins_ltt9779}. In this scenario the star's tidal bulge lags behind the tidal bulge of the planet, generating a net torque that spins up the star and causes the planet to spiral inwards \citep{levrard_falling_2009,matsumura_tidal_2010}. For a given arrangement, the rate of the decay of the orbital period is dependent on the efficiency of tidal energy dissipation within the star, parameterised by the modified stellar tidal quality factor $Q'_*$ \citep{penev_empirical_2018}. Directly measuring the orbital decay rate via long-term transit timing measurements provides an estimate of $Q'_*$, with a larger value being associated with less efficient dissipation and therefore a slower decay rate. While many studies have searched for evidence of orbital decay \citep[e.g.,][]{patra_continuing_2020,hagey_etd}, the only highly convincing detection so far has been for WASP-12\,b \citep[e.g.,][]{Macie_2016,patra_apparently_2017,yee_orbit_2019,turner_decaying_2020,wong_w12}.

Therefore, in addition to fitting the timing data with a constant-period model, we fit an orbital decay model to the mid-times and radial velocity data of LTT\,9779\,b. For a planet on a Keplerian circular orbit with a constant orbital period $P$, we expect the central time of the transits $t_{tra}$ and the occultations $t_{occ}$ to increase linearly. Therefore, these can be described by,
\begin{eqnarray}
t_{tra} &=& t_0 + PE\rm~and\\ 
t_{occ} &=& t_0 + \frac{P}{2} + PE\rm,
\label{equation:linear}
\end{eqnarray}
where $t_0$ is the mid-transit time of the reference epoch and $E$ is the orbit number. In the case of orbital decay, the gradual shrinking of the orbital period gives rise to nonlinear drift in the observable transit and eclipse mid times. The simplest way to represent this behaviour is to include a quadratic term in the expected transit and eclipse centre times. The model is given by,
\begin{eqnarray}
t_{tra} &=& t_0 + PE + \frac{1}{2}\frac{dP}{dE}E^2\rm~and\\
t_{occ} &=& t_0 + \frac{P}{2} + PE + \frac{1}{2}\frac{dP}{dE}E^2\rm.
\label{equation:decay}
\end{eqnarray}
where $\frac{dP}{dE}$ is the period derivative, which we assume to be constant. If negative, it signifies that the orbit of the planet is decaying. We again sample the parameter space using \texttt{Multinest} and compare the Bayesian evidence between each fit to determine the preferred model. The constant period model has two free parameters: the reference epoch $t_0$ and the period $P$. The decay model has three parameters, adding the $dP/dE$ term. As we fit the mid-times and radial velocity data simultaneously, each model has the additional RV model parameters of the velocity semi-amplitude ($K$) and systemic velocities of HARPS (V$_{\rm H}$) and Coralie (V$_{\rm C}$). 

On the other hand, if the orbit of LTT\,9779\,b is slightly eccentric, one may expect the argument of periapse to precess over time due to a number of effects (see Section \ref{sec:orbit}) which would induce sinusoidal variations in the transit and eclipse timing. Hence, we also attempt to fit an apsidal precession model to the timing data. In this case, the transit and eclipse times can be expressed as:

\begin{eqnarray}
t_{tra} &=& t_0 + P_s E - \frac{e P_a}{\pi}\cos{\omega}\rm, \\
t_{occ} &=& t_0 + \frac{P_a}{2} + P_s E +  \frac{eP_a}{\pi}\cos{\omega}\rm,\\
\omega(E) &=& \omega_0 + \frac{d\omega}{dE}E\rm,~and \\
P_s &=& P_a\left(1-\frac{d\omega/{dE}}{2\pi}\right)\rm,
\label{equation:precession}
\end{eqnarray}

for argument of pericentre $\omega$, phase $\omega_0$, and precession rate $d\omega/dE$. In these equations, $P_s$ represents the planet's sidereal period, which is assumed to be fixed, while $P_a$ is the ``anomalistic'' period. This latter period is also fixed, but accounts for the additional period signal due to precession. We note that this approach is an approximation and is only valid for e$<<$0.1 \citep{ragozzine_probing_2009}, but, given the very short period of LTT\,9779\,b and an upper limit of $e<0.058$ at 95\% confidence from \citet{jenkins_ltt9779}, this assumption is valid in this scenario. For the precession model there are a total of eight free parameters ($t_0$, $P$, $\omega_0$, $d\omega/dE$, $e$, $K$, V$_{\rm H}$, \& V$_{\rm C}$).

\section{Results}

\subsection{Atmospheric Composition}
\label{sec:res_atm_ret}

\begin{figure}
    \centering
    \includegraphics[width=\columnwidth]{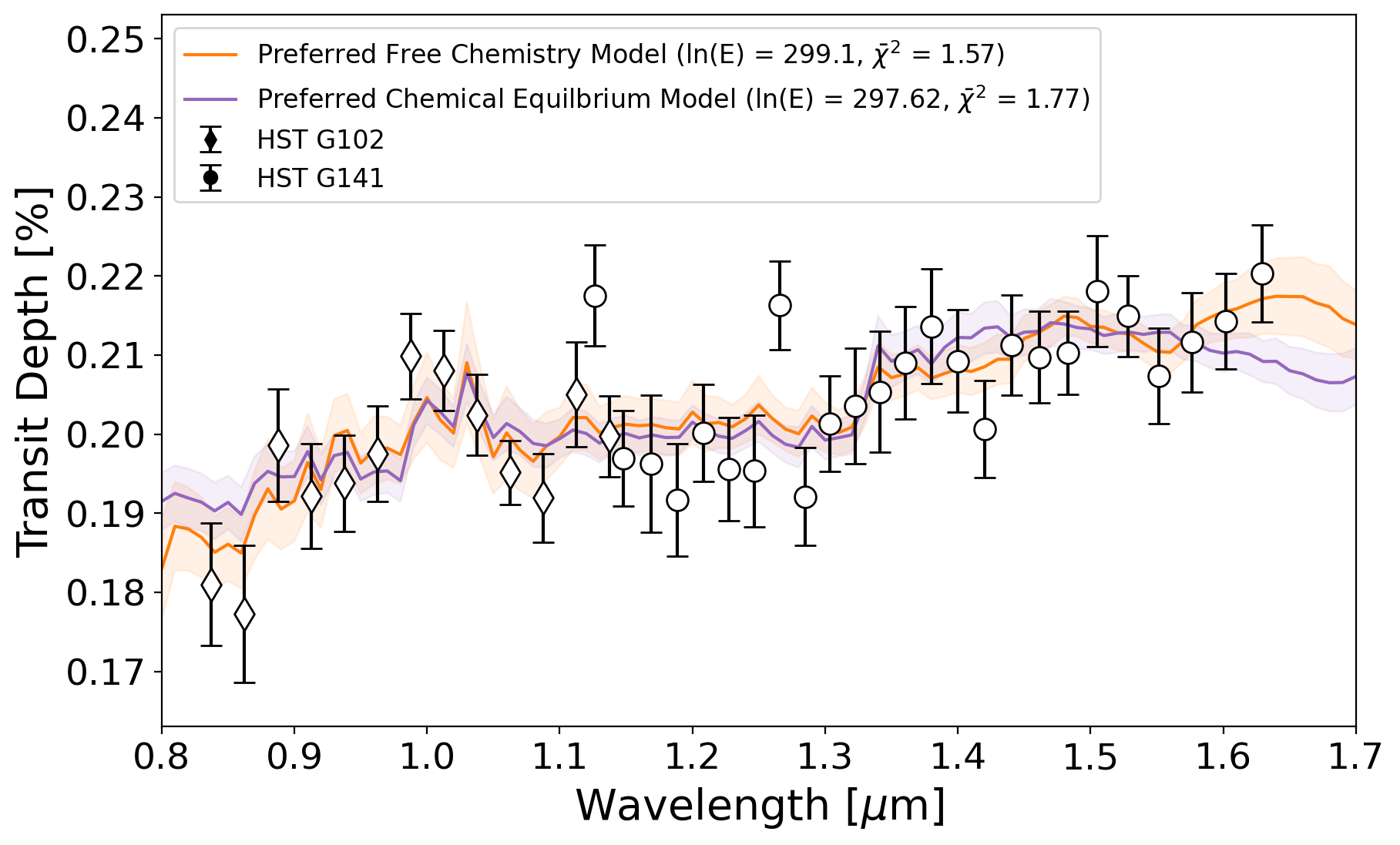}
    \caption{The obtained spectrum from HST WFC3 G102 and G141. Over-plotted are the preferred free chemistry (orange) and chemical equilibrium (purple) models. Based on the Bayesian evidence, the free chemistry model is preferred to 2.4 $\sigma$.}
    \label{fig:spectra}
\end{figure}

We conduct a number of atmospheric retrievals using \texttt{TauREx 3.1} using both free chemistry and chemical equilibrium models. Based on the Bayesian evidence, the preferred model is a free chemistry retrieval which assumed an isothermal atmosphere and for abundances of H$_2$O and CO$_2$ that changed with altitude. The model is preferred to that with constant chemical abundances by 2.41 $\sigma$. The preference between this free model and the chemical equilibrium retrieval with a non-isothermal temperature profile is 2.40 $\sigma$. The non-isothermal chemical equilibrium retrieval is preferred to its isothermal counterpart by 2.48 $\sigma$. We show the best-fit spectra for the preferred free and chemical equilibrium models in Figure \ref{fig:spectra}.

\begin{figure*}
    \centering
    \includegraphics[width=\textwidth]{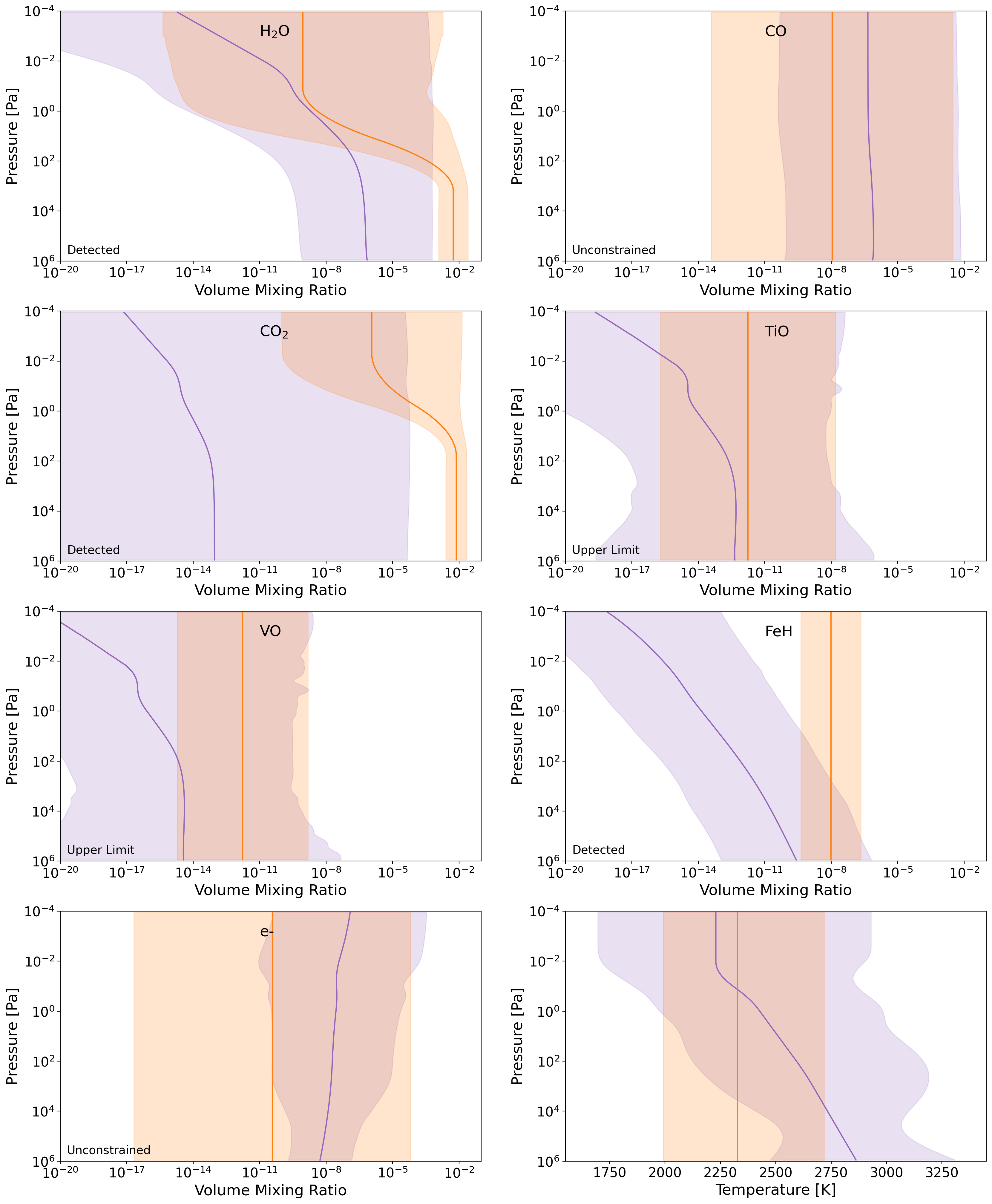}
    \caption{Retrieved molecular abundances from free chemistry (orange) and chemical equilibrium (purple) retrievals. Our free chemistry retrievals suggest the presence of H$_2$O, CO$_2$ and FeH. They also place upper limits on the abundances of TiO and VO but could not constrain CO or e-. The preferred abundance of CO$_2$ is very different between the models, as is the H$_2$O abundance in the lower atmosphere. The retrieved temperature profiles also slightly differ, but are generally within 1 $\sigma$.}
    \label{fig:ret_abundances}
\end{figure*}

Our preferred free chemistry retrieval favour the presence of H$_2$O, CO$_2$ and FeH while also placing upper limits of the presence of TiO and VO. The retrieved abundances for our preferred free and chemical equilibrium retrievals are compared in Figure \ref{fig:ret_abundances}. While many of the abundances agree across these two models, the CO$_2$ abundance is far higher in the free chemistry case, with the H$_2$O abundance in the lower half of the atmosphere also being noticeably higher. While our models preferred the presence of CO$_2$, WFC3 data only has limited sensitivity to carbon-bearing species and this can mean that the molecules retrieved is very dependent upon how the retrieval is setup \citep[e.g.,][]{changeat_k11}.

Our preferred chemical equilibrium retrieval is that which uses a non-isothermal profile. It is likely that this allows for the model to have more flexibility in the dissociation of molecules with altitude compared to when an isothermal profile is used. However, when comparing the retrieved metallicity and C/O ratio we find that the temperature profile did not overly impact the results. We note again that the HST WFC3 range does not give a strong sensitivity to carbon-bearing species and so the C/O ratio is not only poorly constrained but also not likely to be trustworthy \citep{rocchetto}.

Spectral data from different instruments and epochs cannot be automatically assumed to be compatible. Recovering absolute transit depths is extremely difficult and the corner plots of the white light curve fit for the HST WFC3 G141 observations (Figure \ref{fig:white_lc}) show a clear correlation between the transit depth and the systematics model. Therefore, despite the datasets seemingly being compatible when plotted together, we also perform retrievals which allow a global offset to be applied to the G141 data. For these, we utilise the setups from our preferred free chemistry and chemical equilibrium retrievals, with the only change being the addition of this offset parameter.

\begin{figure}
    \centering
    \includegraphics[width=\columnwidth]{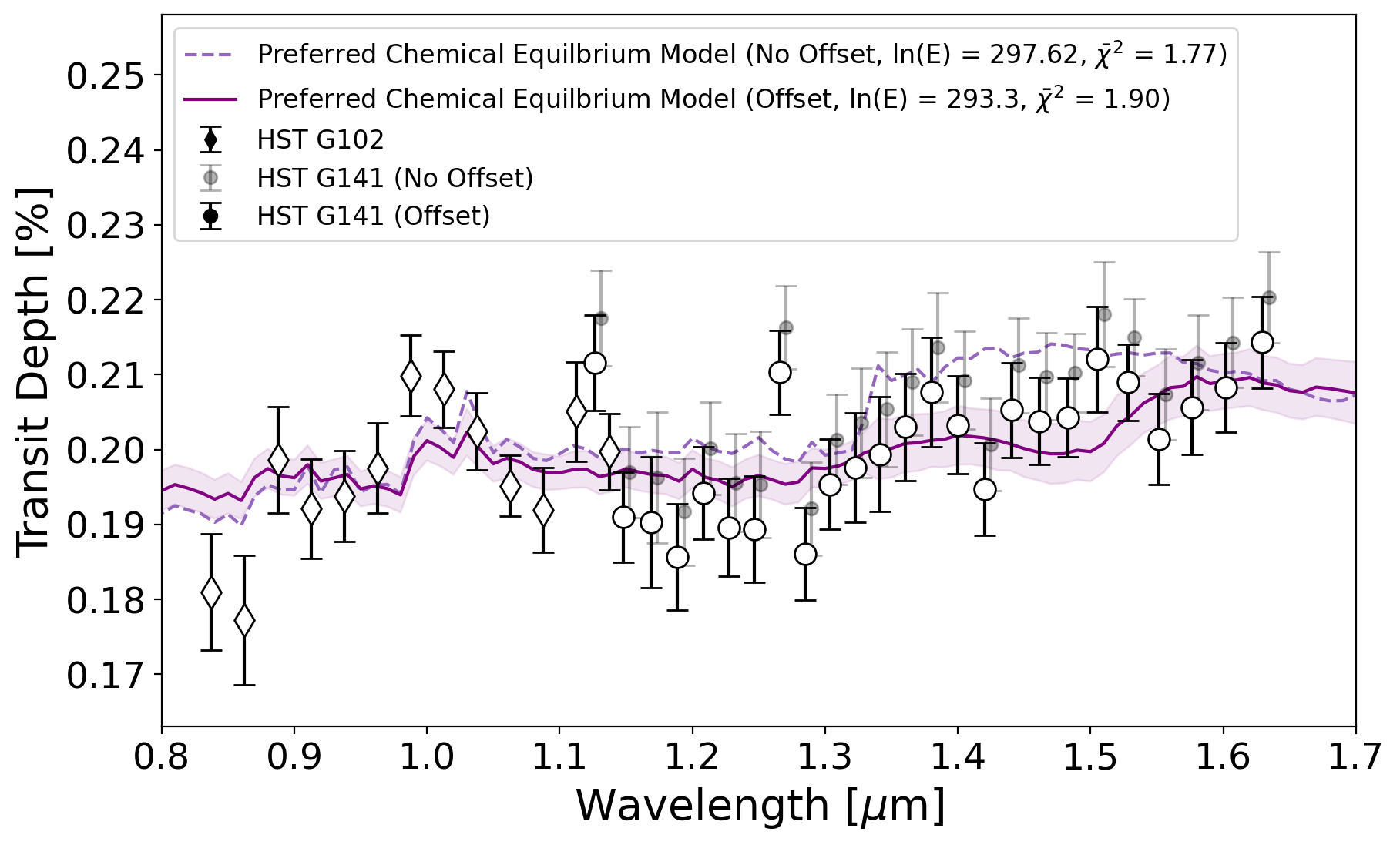}
    \caption{Spectrum from the chemical equilibrium retrieval which included an offset for the HST WFC3 G141 data. The retrieval preferred to reduce the depth of the G141 data by 60 ppm but the solution is not preferred over a retrieval without an offset.}
    \label{fig:spectra_offset}
\end{figure}

We find that, in the free chemistry case, the retrieval converges to a solution with a slight offset. However, given the 1 $\sigma$ uncertainties on this parameter, it is also consistent with there being no offset (-10$^{+33}_{-38}$ ppm). Meanwhile, the chemical equilibrium retrieval prefers a more significant offset (-60$\pm$26 ppm) and we show the preferred solution, including the offset to the HST WFC3 G141 data, in Figure \ref{fig:spectra_offset}. However, in both cases, the Bayesian evidence shows that the models without offsets were preferred to 3.96 and 3.4 $\sigma$ for the free and chemical equilibrium cases respectively. 

\begin{figure}
    \centering
    \includegraphics[width=\columnwidth]{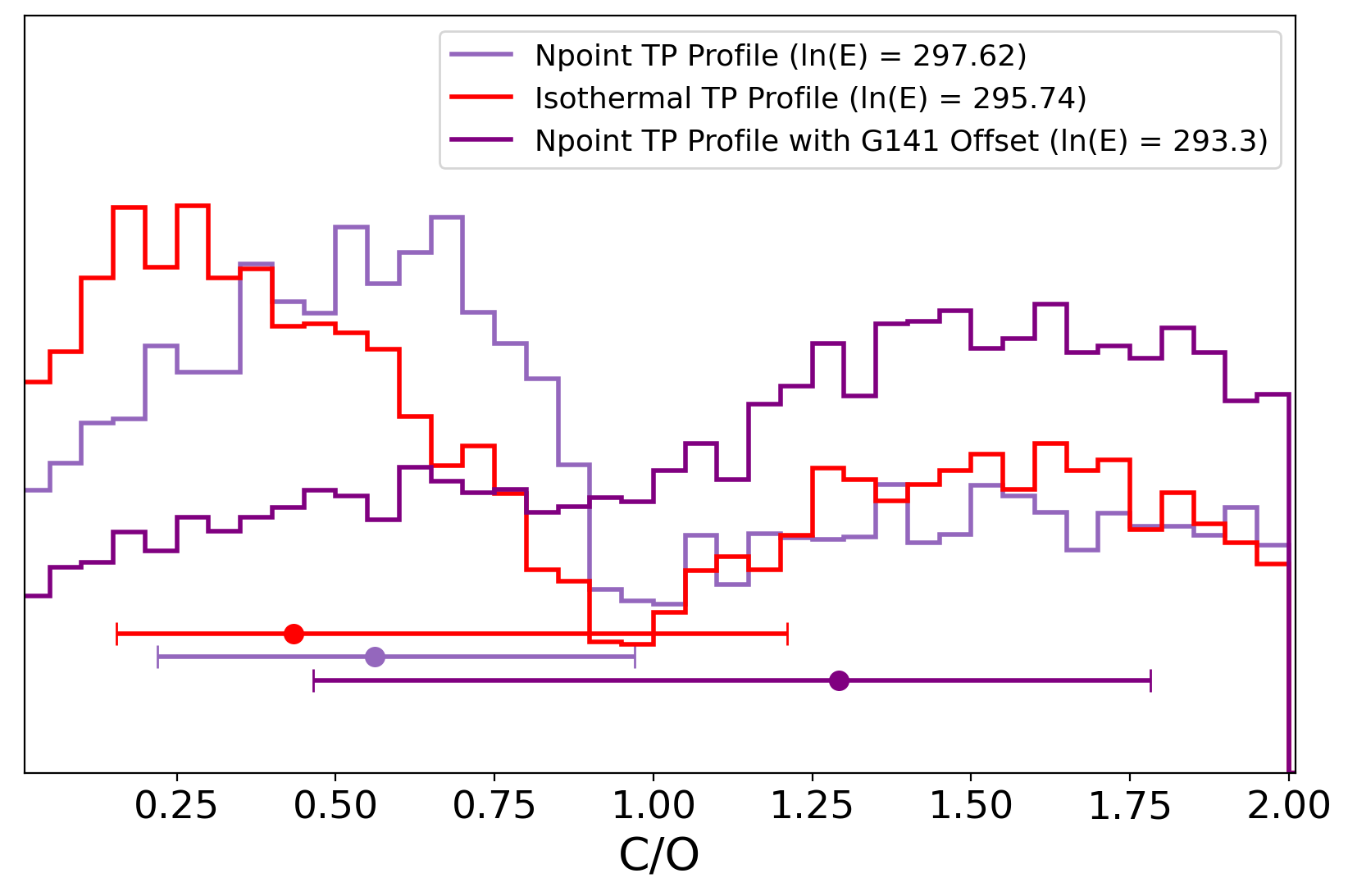}
    \includegraphics[width=\columnwidth]{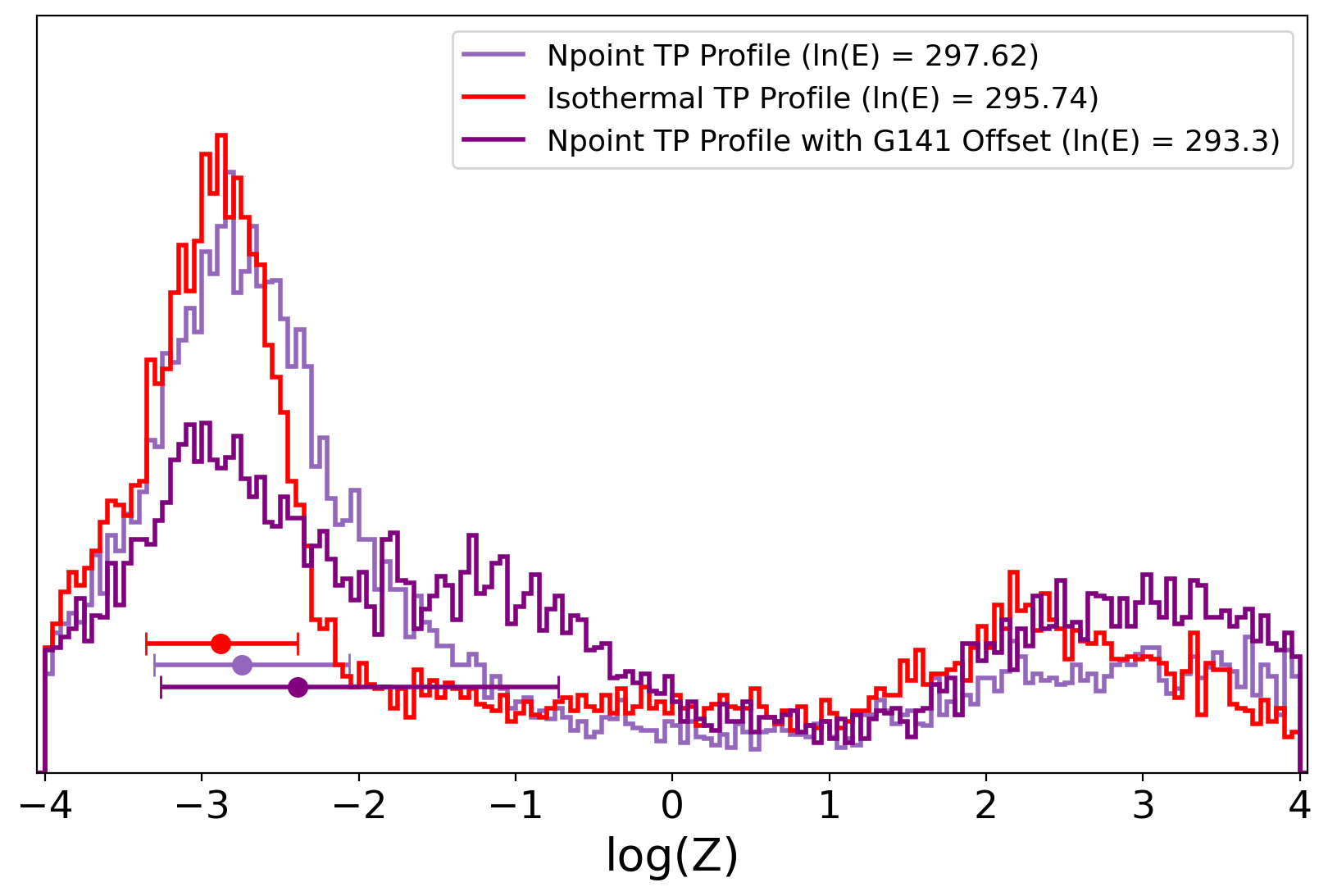}
    \caption{Probability distributions for the carbon-to-oxygen ratio (C/O, top) and atmospheric metallicity (log$_{10}$(Z), bottom) from the different chemical equilibrium retrievals conducted here. The constraints placed on these proprieties are poor in each case, but all three prefer solution with very low atmospheric metallicities.}
    \label{fig:chem_eq_res}
\end{figure}

Due to the slope of the spectrum, we investigate the possibility that unocculted spots and faculae are the cause of the modulation seen. Unocculted faculae in particular are capable of reproducing the observed negative bluewards slope. Our free-chemistry retrievals in which this effect was accounted for lead to a fit which has a highly similar Bayesian evidence to that of the free-chemistry model alone but requires 4 additional fitting parameters. The inclusion of the modelling of stellar heterogeneities removes any strong constraints on the atmosphere, with no molecules being conclusively detected. However, the retrieval converges to an extremely high faculae filling factor (65\% $\pm$ 12 of the unocculted stellar surface). Given that the star is known to be slowly rotating and therefore inactive \citep{jenkins_ltt9779}, and that there have been no indications of occulted faculae in any transit observations taken of the LTT\,9779\,b, this value is likely unrealistic.

We provide an overview of our retrieval setups, as well as the associated evidence and reduced chi-square values, in Table \ref{tab:ret_results}.

\subsection{Search for the 1.083 $\mu$m Helium Line}

\begin{figure}
    \centering
    \includegraphics[width=\columnwidth]{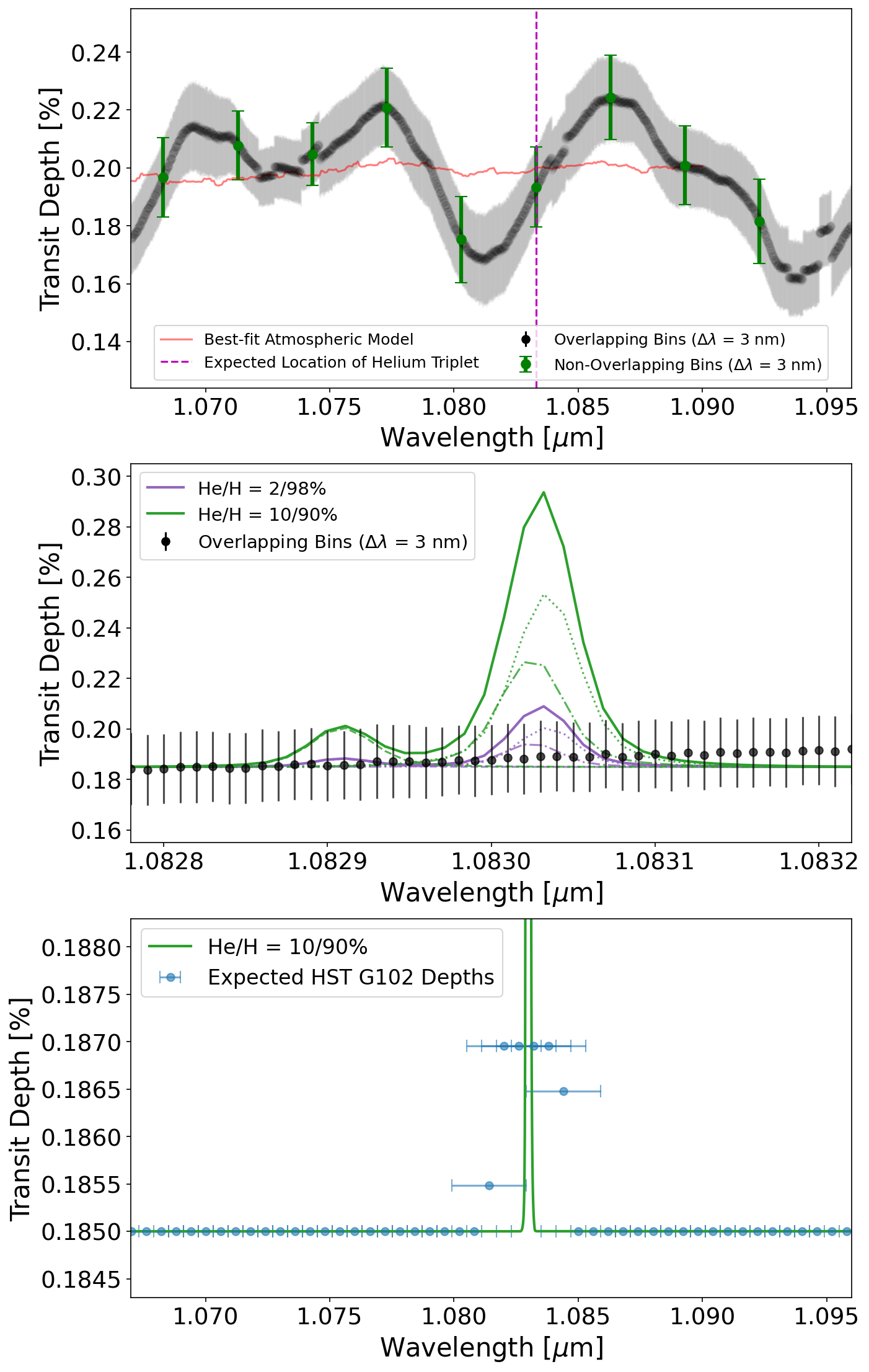}
    \caption{Search for the 1.083 $\mu$m Helium triplet. While it is expected that the planet is undergoing mass loss, little evidence could be found via this tracer with the data collected with HST WFC3 G102. \textbf{Top:} the higher-resolution HST WFC3 G102 spectrum extracted in this work. The bins are spaced by 0.1 \r{A} but have a bandwidth of 3 nm and, therefore, are overlapping. The location of the peak of the Helium triplet signature is indicated by the purple dashed line. The best-fit atmospheric model from Section \ref{sec:res_atm_ret} is also shown (red) for context, as are a set of bins which are not overlapping (green). \textbf{Middle}: expected signal of the Helium triplet derived from simulations of atmospheric escape driven by photoevaporation (Allan et al, submm.). The line is formed of three segments (dotted, dashed, dot-dash) which together produce the final signal (solid). The expected triplet signature is shown for two He/H ratios. The HST WFC3 G102 data is also shown, with no excess absorption seen in these bins. \textbf{Bottom:} our simulated Helium signature when convolved with the bins extracted. As the spectral bins are overlapping, the expected signal in these bins from the triplet appears as a plateau. The size of the signal ($\sim$20 ppm) is much smaller than the uncertainties on each data point ($\sim$ 135 ppm), even in the 10/90 He/H case. Thus we conclude that the data is not sensitive enough to detect the expected triplet signature. }
    \label{fig:helium_line}
\end{figure}

As well as the low-resolution spectrum used for our atmospheric retrievals, we also extract a higher-resolution spectrum for the G102 data with the aim of searching for the Helium line at 1.083 $\mu$m that would be indicative of ongoing mass-loss. However, despite sampling this spectral range with overlapping spectral bins which had a spacing of only 0.1 \r{A} and a bandwidth of 3 nm (30 \r{A}), we find no obvious evidence of excess absorption at, or around, this wavelength (see top panel of Figure \ref{fig:helium_line}). The variations that are seen in this spectrum are likely due to correlated noise.

We also model the predicted atmospheric evaporation using the methods of \citet{allan2019} and the reconstructed high-energy spectrum of LTT\,9779 presented in Section \ref{meth:he_line}. We find that assuming either a 2/98\% or 10/90\% helium to hydrogen number abundance, the evaporation rate is expected to be $\sim10^{11}$ g/s, reaching velocities of up to $\sim 50$~km/s. However, even though the planet is anticipated to have a substantial evaporation rate driven by photoionisation, the expected signature of the evaporation in the triplet lines is very small, particularly in the 2/98\% case. The middle panels of Figure \ref{fig:helium_line} show the anticipated transmission spectrum in the helium triplet, where we derive a peak transit depth of 0.29\%. Due to the width of the spectral bins (3 nm/30 \r{A}), the Helium triplet would not appear as a sharp peak but as a plateau in the data: multiple bins would completely contain the excess absorption. In the bottom panel of Figure \ref{fig:helium_line} we show the magnitude of this plateau when our 10/90\% He/H model is convolved with these spectral bins. Even in the 10/90\% case, the expected signature strength is far smaller than the uncertainties on the G102 spectra data ($\sim$20 ppm versus $\sim$135 ppm). Therefore, we conclude the data is not sufficient to detect this tracer as even though we expect LTT\,9779\,b has a significant evaporation rate, its evaporation is barely seen in the helium triplet. 


\subsection{Updated Ephemeris and Search for Long-Term Timing Variations}\label{sec:updated_ephemeris}

\begin{figure}
    \centering
    \includegraphics[width=0.9\columnwidth]{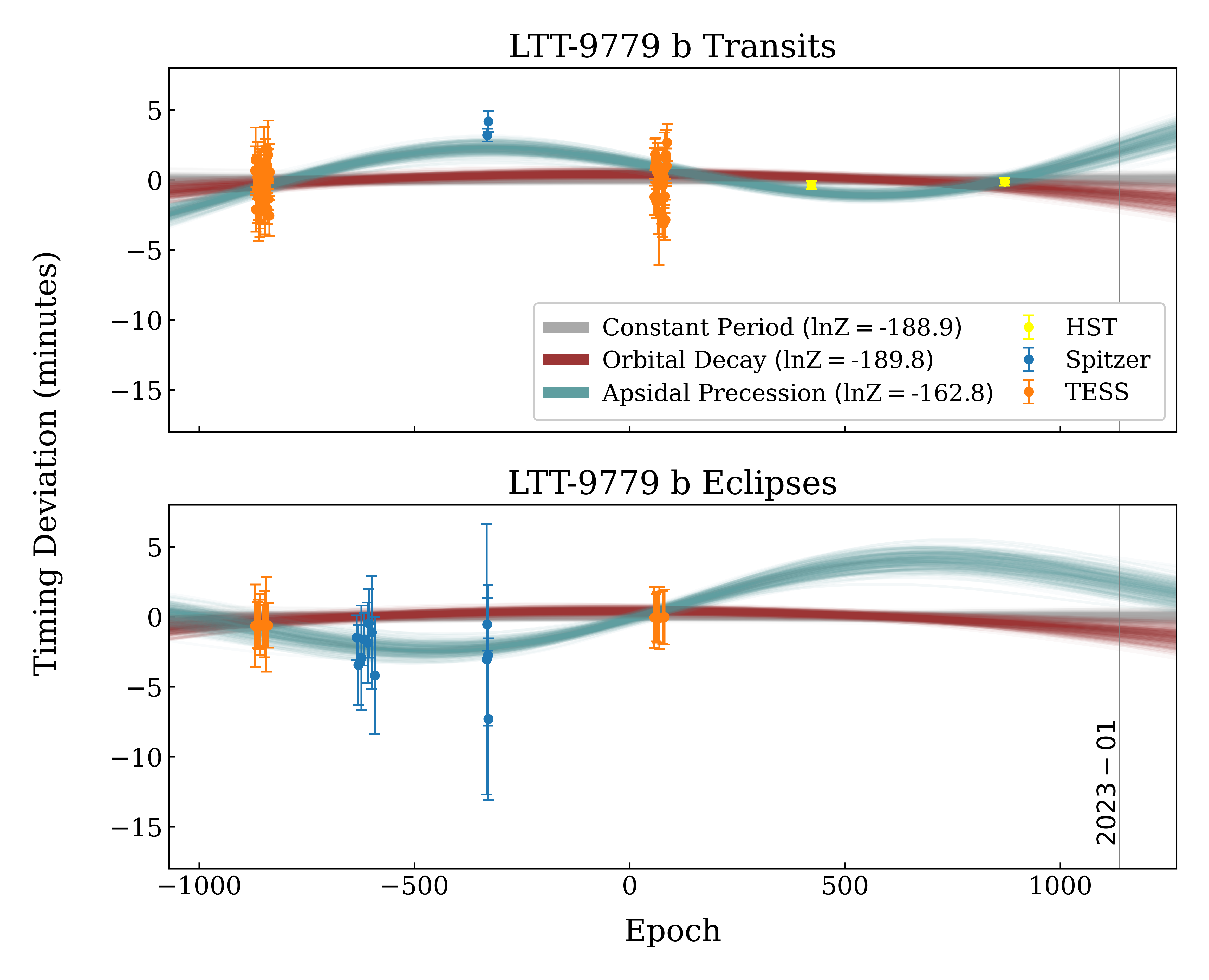}
    \includegraphics[width=0.9\columnwidth]{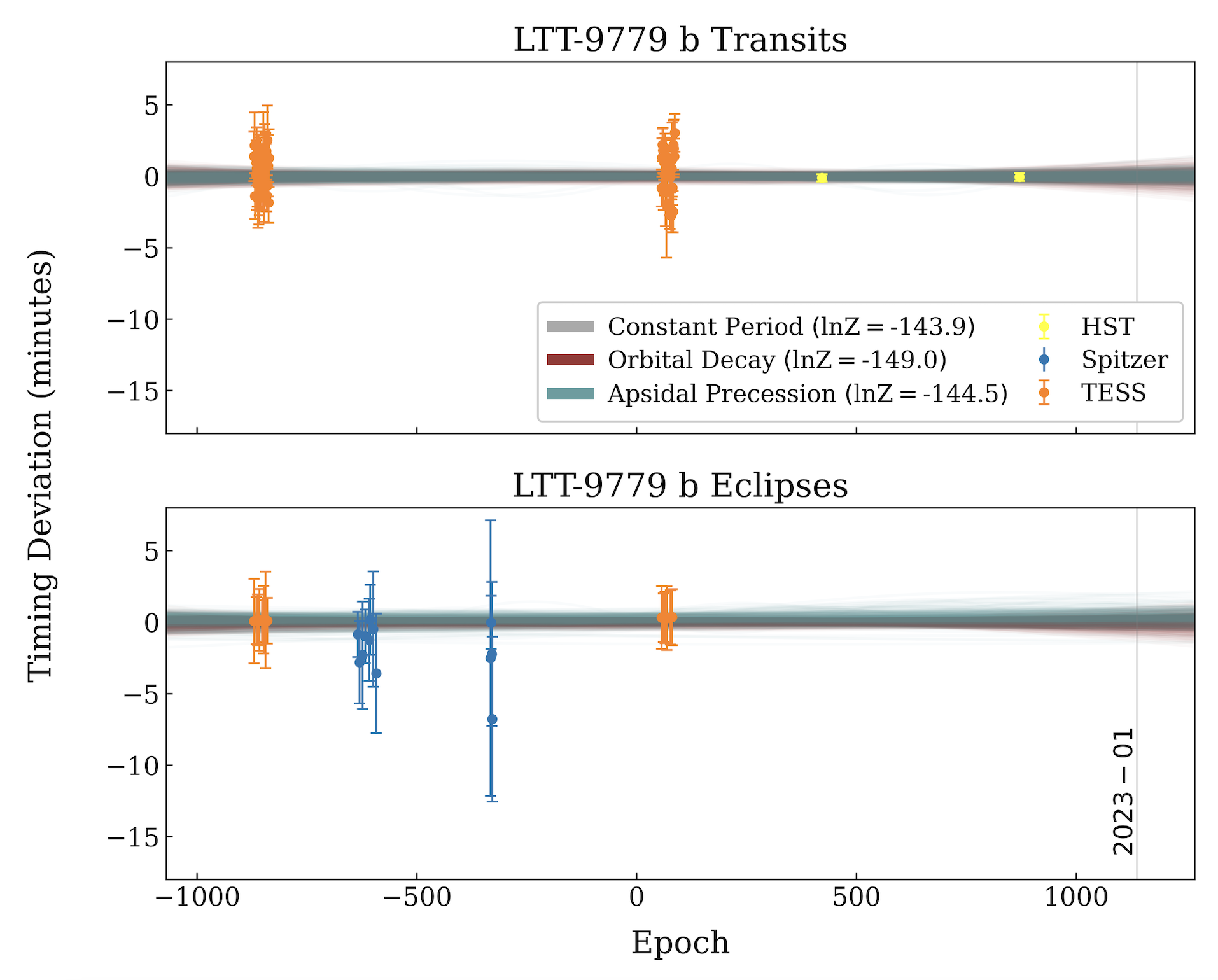}
    \includegraphics[width=0.9\columnwidth]{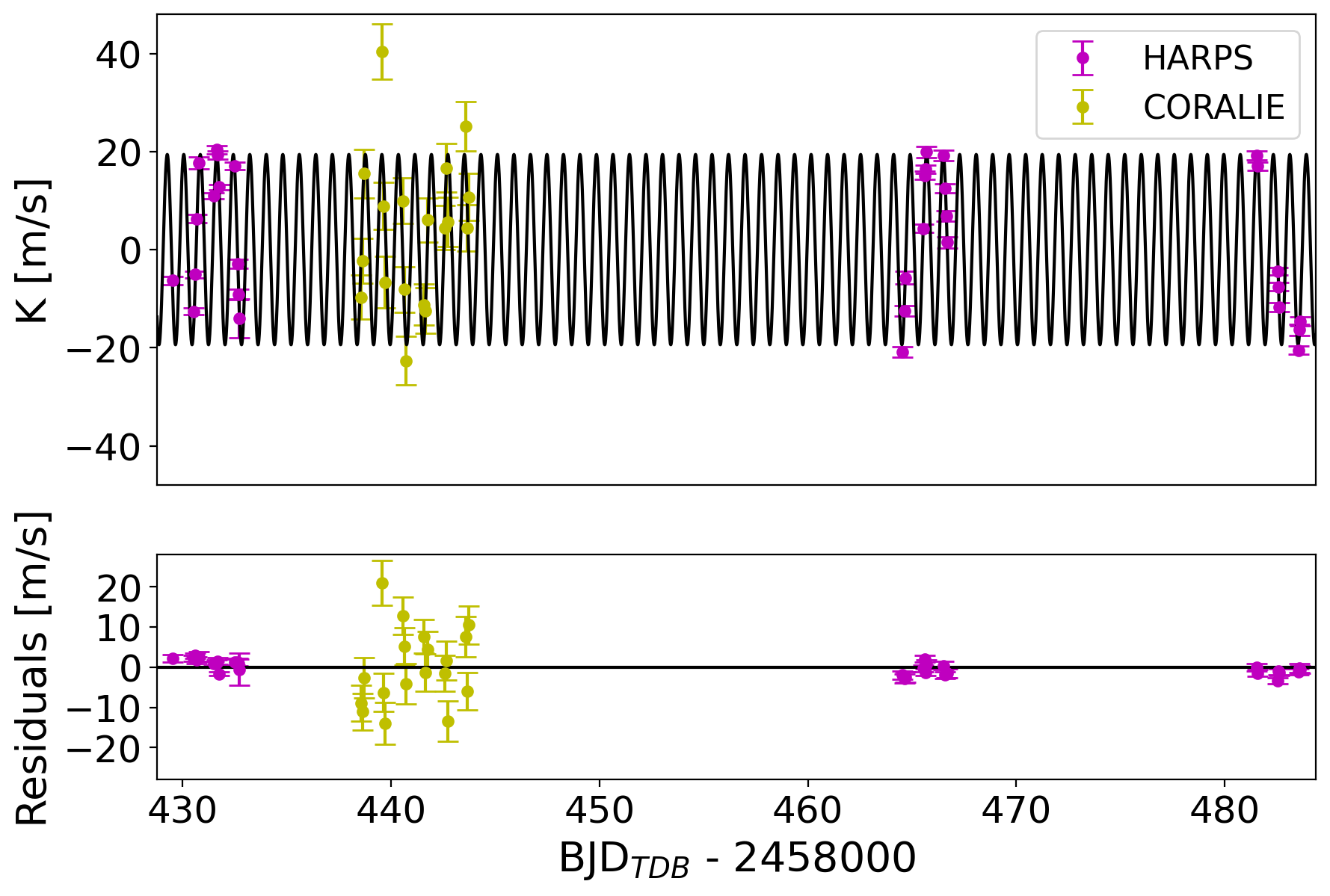}
    \caption{Plots from our ephemeris fitting for LTT\,9779\,b. \textbf{Top:} Observed-minus-calculated (O-C) plots for all transit and eclipse mid-times. The reference epoch is set to be halfway between the first and last transit used in this work. Traces from all three models, linear (grey), decay (red), and precession (turquoise), are also shown. \textbf{Middle:} O-C plots once the Spitzer transits have been discarded. In doing so, all evidence for non-linear ephemerides is also removed. \textbf{Bottom:} fit to the literature radial velocity data. For brevity, we only show the best-fit model to this data when considering a constant period and when the Spitzer transit mid-times have been discarded, which we adopt as our favoured model. However, these data are used in all our fits.}
    \label{fig:oc}
\end{figure}

We fit the transit and eclipse mid-times with three models: linear ephemeris, orbital decay, and apsidal precession. We find no statistical evidence to prefer orbital decay over a linear ephemeris: the linear model has a Bayesian log evidence of ln(E) = -225.1 while the decay yielded ln(E) = -224.9. The apsidal precession model, however, is highly favoured with a Bayesian evidence of ln(E) = -201.2. Examination of the data shows that the Spitzer transits have a significant observed minus calculated (O-C) residual compared to the uncertainties on the mid-time (see Figure \ref{fig:oc}). These were taken from fits conducted by \citet{crossfield_ltt} to the phase curves of LTT\,9779\,b. The eclipse mid-times from these phase curves \citep[which were taken from the same data but independently fitted by, ][]{dragomir_ltt} do not show, in contrast, the same magnitude of deviation, thereby suggesting these points may be outliers rather than an actual detection of a significant deviation from the expected time.

Due to the shape of the decay and precession models within an O-C plot (i.e., a parabola and a sin curve), we surmise that these Spitzer transit mid-times could be driving the model towards a solution that favours precession, with the large gaps in the transit coverage also enabling such a solution. Therefore, we also explore fits to the data without the Spitzer transit mid-times. In this case, the linear model is preferred with ln(E) = -181.0 versus ln(E) = -185.8 and ln(E) = -181.7 for the orbital decay and apsidal precession models, respectively. Though the evidence against apsidal precession versus a linear ephemeris is only marginal, the best-fit solution for the precession model is linear within the uncertainty (see Table \ref{tab:lin_dec_pre_mod}). The inclusion or exclusion of the Spitzer transit mid-times has a minimal impact on the best-fit linear period but, given the potentially spurious nature of the mid-times, we report the recovered period without these points in Table \ref{tab:para}. For completeness, the best-fit parameters for all fits are given in the Appendix.

\section{Discussion}

\subsection{The Atmosphere of LTT\,9779\,b}

\begin{figure}
    \centering
    \includegraphics[width=\columnwidth]{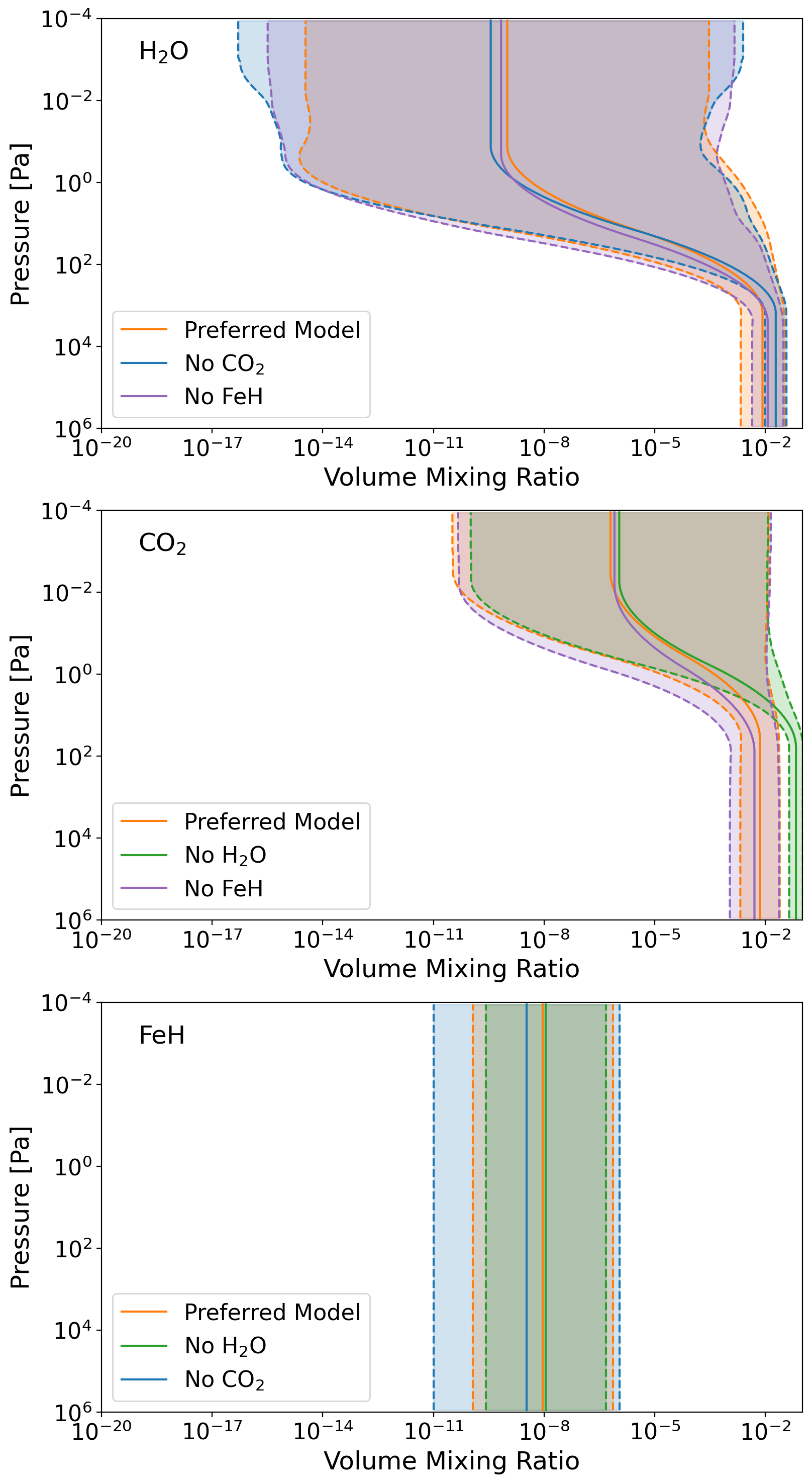}
    \caption{Changes to the retrieved abundances of H$_2$O, CO$_2$ and FeH when one of the other molecules is removed from the retrieval. In all three cases, the model without the molecule is not preferred to the full setup. Comparing the Bayesian evidence suggests the presence of H$_2$O is preferred to 3.11 $\sigma$ while there is a 2.35 $\sigma$ preference for the inclusion of CO$_2$ and a 2.12 $\sigma$ preference for FeH. All these retrievals were conducted with an isothermal temperature-pressure profile.}
    \label{fig:alt_ret}
\end{figure}

We have presented transmission observations of the hot Neptune LTT\,9779\,b, extracting a spectrum from 0.8-1.6 $\mu$m. Our best-fit free chemistry model shows evidence for high abundances of H$_2$O and CO$_2$. To test the strength of each of these detections, we run additional retrievals. Without H$_2$O in the model, the best-fit model provides a worse fit to the data with the Bayesian evidence suggesting that the presence of H$_2$O is preferred to 3.12 $\sigma$. In this model without H$_2$O, the retrieved CO$_2$ abundance is even higher than in our preferred model as shown in Figure \ref{fig:alt_ret}. We also run a retrieval without CO$_2$, finding that the Bayesian evidence was reduced in this case: the inclusion of CO$_2$ was preferred to 2.01 $\sigma$. The lower significance of this detection is likely due to the fact that WFC3 data only has limited sensitivity to carbon-bearing species. Furthermore, it has been shown that carbon-bearing species identified within a preferred model can be dependent upon how the retrieval is setup \citep[e.g.,][]{changeat_k11}. Interestingly, the removal of CO$_2$ from the model does not lead to the detection of CO or HCN. Instead, the H$_2$O abundance is affected with a slightly higher volume mixing ratio being preferred (see Figure \ref{fig:alt_ret}). The other molecular species that was constrained in the free-chemistry case was FeH and we found that our retrievals preferred its presence to 2.13 $\sigma$. Hence, for all these species we find a relatively weak detection significance.

\begin{figure}
    \centering
    \includegraphics[width=\columnwidth]{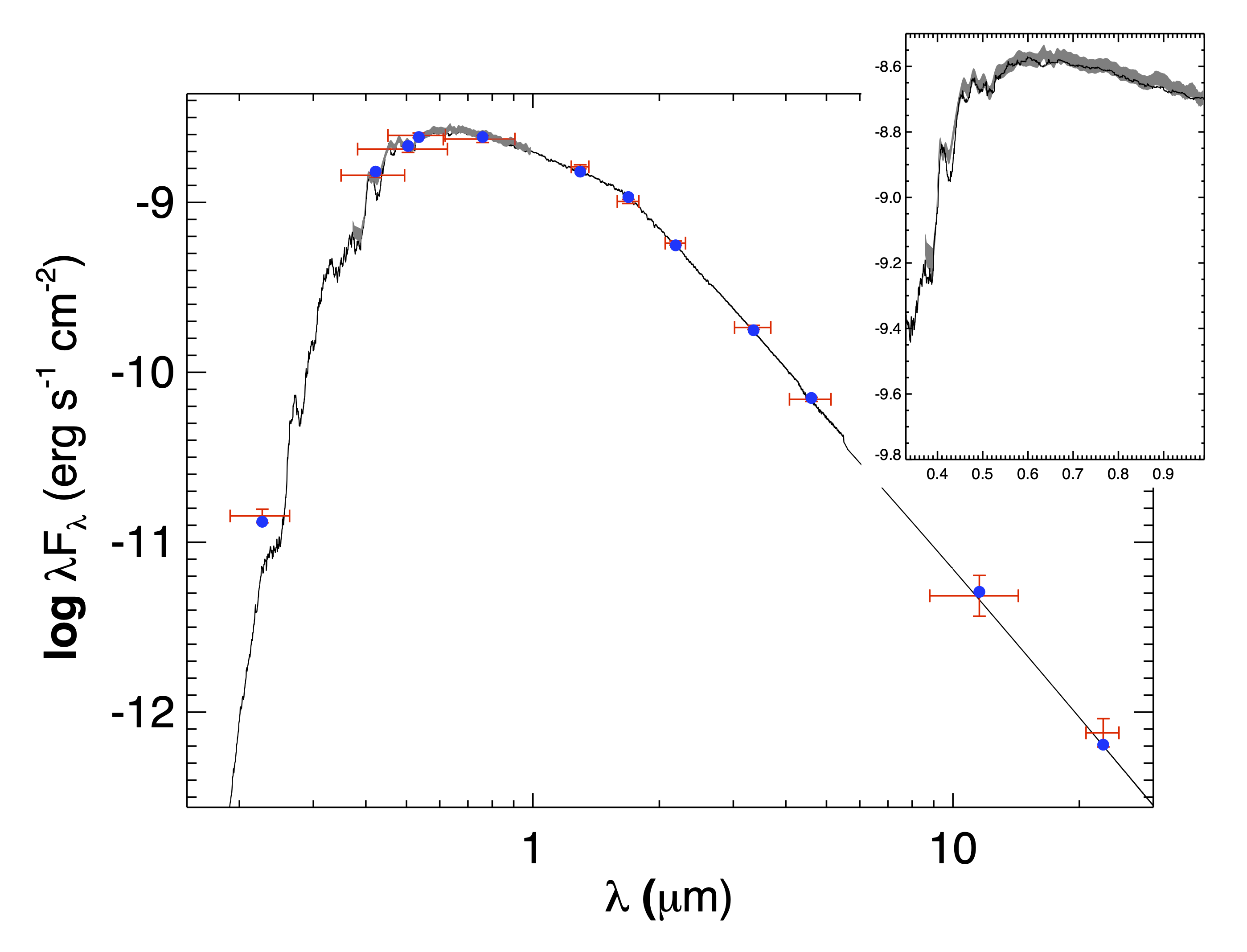}
    \caption{Spectral energy distribution of LTT 9779. Red symbols represent the observed photometric measurements, where the horizontal bars represent the effective width of the passband. Blue symbols are the model fluxes from the best-fit \texttt{PHOENIX} atmosphere model (black). The {\it Gaia\/} spectrum is also shown as a grey swathe; the inset plot shows a close-up.}
    \label{fig:sed}
\end{figure}

Our retrievals with \texttt{ASteRA} prefer models with a high spot coverage. Such coverages are unrealistic, but could potentially be caused by assuming the wrong star temperature in our models. As the spots in the model are hotter than the rest of the stellar surface, the unrealistic coverage might indicate that our assumed stellar temperature is too low. We explore this possibility by independently determining the basic stellar parameters. We perform an analysis of the broadband spectral energy distribution (SED) of the star together with the {\it Gaia\/} DR3 parallax \citep[with no systematic offset applied; see, e.g.,][]{stassun_torres}, in order to determine an empirical measurement of the stellar radius, following the procedures described in \citet{stassun_2016,stassun_2017,stassun_2018}. We pull the $B_T V_T$ magnitudes from {\it Tycho-2}, the $JHK_S$ magnitudes from {\it 2MASS}, the W1--W4 magnitudes from {\it WISE}, the $G_{\rm BP} G_{\rm RP}$ magnitudes from {\it Gaia}, as well as the NUV flux from {\it GALEX}. Together, the available photometry spans the full stellar SED over the wavelength range 0.2--20~$\mu$m (see Figure~\ref{fig:sed}).  
 
We perform a fit using \texttt{PHOENIX} stellar atmosphere models, with the free parameters being the effective temperature ($T_{\rm eff}$) and metallicity ([Fe/H]); we set the extinction $A_V \equiv 0$, due to the close proximity of the system, and we assume a surface gravity $\log g \approx 4.5$ as appropriate for a main-sequence dwarf. Finally, we use the {\it Gaia\/} spectrum as a consistency check for the overall absolute flux calibration. 

The resulting fit (Figure~\ref{fig:sed}) has a best-fit $T_{\rm eff} = 5450 \pm 75$~K and [Fe/H] = $0.0 \pm 0.3$, with a reduced $\chi^2$ of 1.4. Integrating the model SED gives the bolometric flux at Earth, $F_{\rm bol} = 3.503 \pm 0.041 \times 10^{-9}$ erg~s$^{-1}$~cm$^{-2}$. Taking the $F_{\rm bol}$ and together with the {\it Gaia\/} parallax directly gives the bolometric luminosity, $L_{\rm bol} = 0.7173 \pm 0.0084$~L$_\odot$, which with the $T_{\rm eff}$ gives the stellar radius, $R_\star = 0.951 \pm 0.018$~R$_\odot$. In addition, we can estimate the stellar mass from the empirical eclipsing-binary based relations of \citet{torres_2010}, giving $M_\star = 0.96 \pm 0.06$~M$_\odot$. Finally, the mass and radius together give the mean stellar density, $\rho_\star = 1.58 \pm 0.13$~g~cm$^{-3}$. 

These values are consistent with those from \citep{jenkins_ltt9779}, which we use for all the analyses here. In particular, the recovered temperature is not higher than assumed: it is, in fact, slightly lower ($T_{\rm eff} = 5450 \pm 75$~K vs $5480 \pm 42$~K). Therefore, it seems that this is not the cause of the high spot coverage preferred in our retrievals with \texttt{ASteRA}.

\begin{table*}[]
    \centering
    \caption{Summary of atmospheric retrievals conducted in this work.}
    \begin{tabular}{cccccccc}\hline\hline
    Chemistry & TP Profile & Molecules Removed$^\dagger$ & 2-layer Molecules & Offset & Spots/Fac & ln(E) & $\bar{\chi}^2$ \\\hline
    Free & Isothermal & None & H$_2$O \& CO$_2$ & False & False & 299.10 & 1.57 \\
    Free & Isothermal & H$_2$O & CO$_2$ & False & False & 295.64 & 1.74 \\
    Free & Isothermal & CO$_2$ & H$_2$O& False & False & 297.51 & 1.62 \\
    Free & Isothermal & FeH & H$_2$O \& CO$_2$ & False & False & 297.93 & 1.73 \\
    Free & Isothermal & None & None & False & False & 297.77 & 1.46 \\
    Free & Isothermal & None & H$_2$O \& CO$_2$ & False & True	& 299.31 & 2.43 \\
    Free & Isothermal & None & H$_2$O \& CO$_2$ & True & False & 293.76 & 1.68 \\
    Free & NPoint & None & H$_2$O \& CO$_2$ & False & False & 298.32 & 2.15 \\
    Free & NPoint & None & H$_2$O \& CO$_2$ & True & False & 294.98 & 2.27 \\
    Free & NPoint & None & None & False & False & 297.61 & 1.87 \\ \hline
    Chem Eq & Isothermal & N/A & N/A & False & False & 295.74 & 1.57 \\
    Chem Eq & NPoint & N/A & N/A & False & False & 297.62 & 1.77 \\
    Chem Eq & NPoint & N/A & N/A & True & False & 293.30 & 1.90 \\ \hline
    \multicolumn{8}{c}{$\dagger$ Compared to standard setup of H$_2$O, CO$_2$, CO, HCN, TiO, VO, FeH, H-}\\\hline\hline
    \end{tabular}
    \label{tab:ret_results}
\end{table*}



\begin{figure}
    \centering
    \includegraphics[width=\columnwidth]{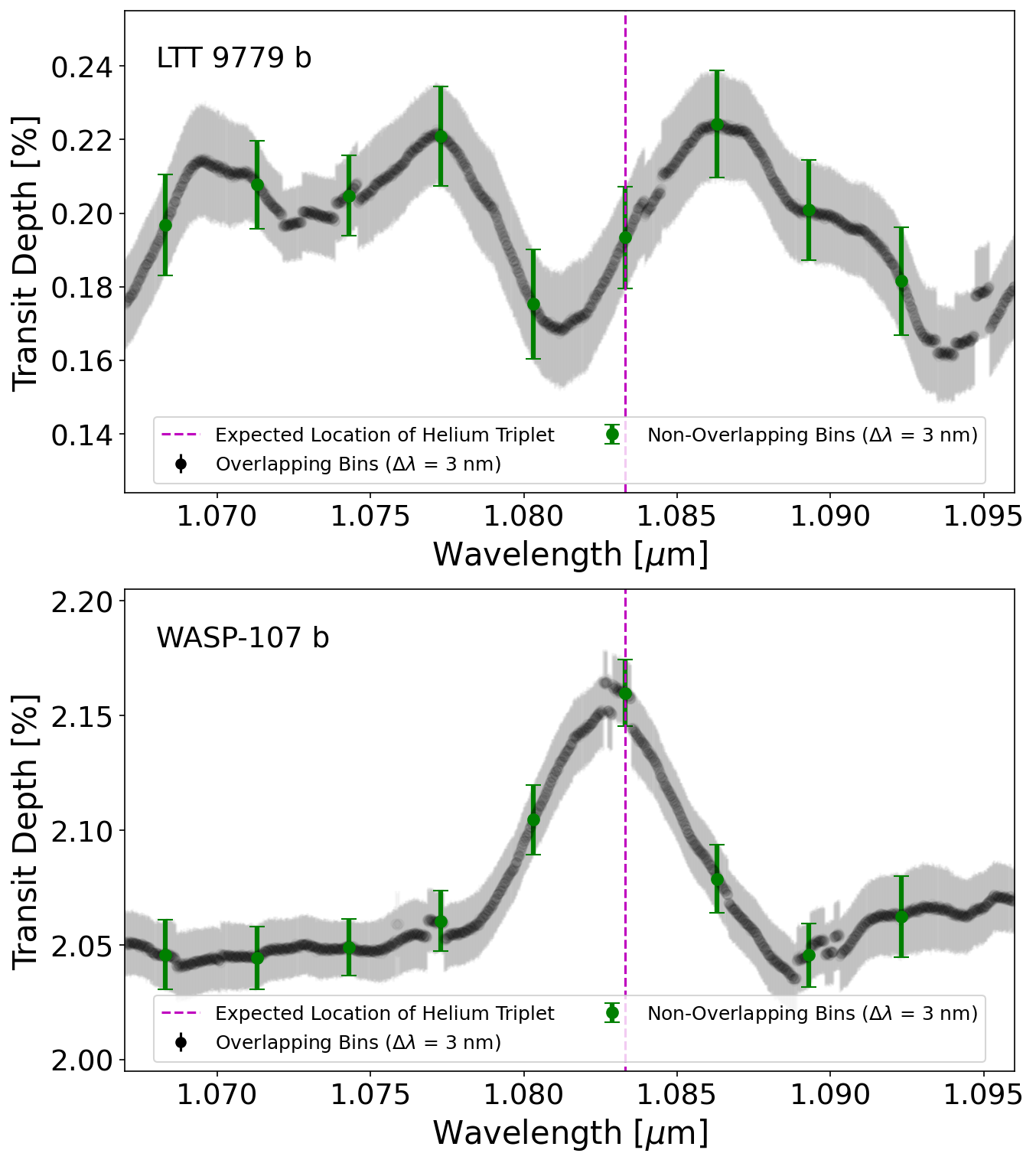}
    \caption{High-resolution ($\delta \lambda$ = 30 \r{A}, spaced by 0.1 \r{A}) HST WFC3 G102 data for LTT\,9779\,b (top) and WASP-107\,b (bottom). In the latter case, the excess absopriotn due to the Helium triplet is clearly apparent.}
    \label{fig:comp_w107}
\end{figure}

Given its high irradiation, it was anticipated that LTT\,9779\,b would be undergoing significant mass loss. Hence, it might be expected that this escape could be observed using tracers of atmospheric escape such as the Helium triplet at 1.083 $\mu$m and thus that the non-detection within the HST WFC3 G102 is surprising. However, it is important to consider that the observability of the atmospheric escape via this particular tracer is not only dependant on the level of atmospheric escape. The population of helium in the triplet state also plays a critical role and this has been shown to depend on the nature of the stellar irradiance. The triplet state is more readily populated for K-dwarf hosts with their relatively low mid-UV flux (which can photoionise helium out of the triplet state) to EUV flux (which is both a significant contributor to driving the escape itself and populating the triplet via recombinations) compared to warmer G-type stars \citep{Oklopcic_2019_dep_st_rad}. Our work shows that LTT\,9779\,b indeed receives a high mid-UV flux, with our model finding that mid-UV photoionisations is the main depopulating source for triplet-state helium and hence largely responsible for lack of a clear detection of Helium at 1.083 $\mu$m. To further cement this non-detection, we apply the same fitting procedure to the HST WFC3 G102 data of WASP-107 b, which is shown in Figure \ref{fig:comp_w107} and highlights the strength of the signal seen in that case. 

\begin{figure}
    \centering
    \includegraphics[width=\columnwidth]{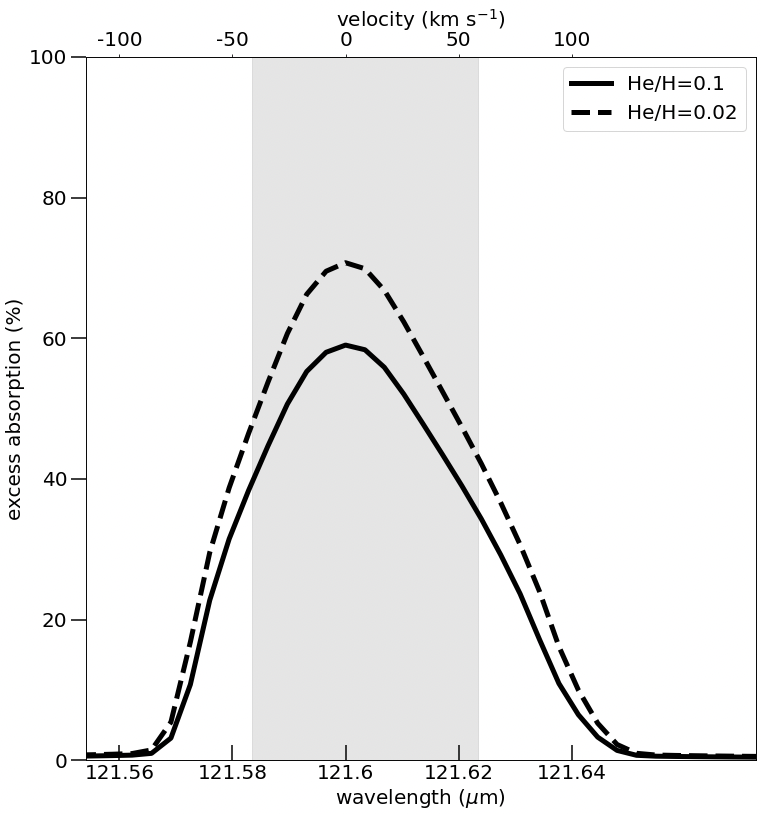}
    \caption{Anticipated Lyman-$\alpha$ signature for LTT\,9779\,b assuming 2/98\% He/H (dotted) and 10/90\% He/H (solid). While no trace of atmospheric escape was found with the Helium triplet, this tracer may offer another chance to detect and/or constrain the escape rate of LTT\,9779\,b's atmosphere.}
    \label{fig:ly_alpha}
\end{figure}

Alternative tracers, such as the commonly used hydrogen Lyman-$\alpha$ line may allow the predicted strong escape of LTT\,9779\,b to be observed. Again utilising the model of \citet{allan2019}, we predict significant absorption in the non-contaminated Lyman-$\alpha$ line wings and show the resulting profile in Figure \ref{fig:ly_alpha}. However, we note that three dimensional modeling considering the interaction with the stellar wind is required for a realistic Lyman-$\alpha$ prediction, given that high velocity blue-shifted absorption originates due to the interaction with the stellar wind \citep{2021MNRAS.500.3382C}. We leave such a detailed analysis for further work.

In their study of Helium escape for planets close to the Neptune desert, \citet{vissapragada_helium} derived a boundary below which a planet could not have lost more than 50\% of its initial mass to photoevaporation. As LTT\,9779\,b is situated below this region, they noted that is was unlikely to be the photoevaporated core of a giant gaseous planet, with high-eccentricity orbital migration being the more likely explanation for the planet's current characteristics as the planet could have been delivered to its current location relatively late. Further evidence for this could present in the planet's atmospheric composition (e.g., a high C/O ratio) or via the detection of a remaining eccentricity i the orbit of LTT\,9779\,b. However, the data analysed here, and in previous studies \citep[e.g.,][]{dragomir_ltt,crossfield_ltt}, is not able to provide such evidence.

Hence, it is hard to conclude a great deal about the nature of the atmosphere of LTT\,9779\,b. We expect that far better constraints on the nature of LTT\,9779\,b's atmosphere from the JWST NIRISS \citep{doyon_niriss} phase-curve observations, which have already been taken as part of Cycle 1. The transit portion of the JWST phase-curve may also shed light on spectral variability when compared to the data presented here. Given the wider spectral coverage and higher SNR, the NIRISS transmission data should also be able to more easily disentangle atmospheric features from stellar contamination. An analysis of the global temperature and chemistry will test the implications made in this study, and those from the Spitzer data, and provide a more accurate interpretation of both. While we do not detect the signature of atmospheric escape from the Helium triplet in this work, the JWST NIRISS observations may provide a detection. Additionally, observations with HST of the Ly$\alpha$ line may provide better constrains on the current escape rate pf LTT\,9779\,b's atmosphere and thus on the nature of the hot Neptune desert.

Furthermore, eclipses observations of LTT\,9779\,b with CHEOPS have found evidence for variability (AO2-013; PI: James Jenkins), with additional observations being taken to confirm this (AO3-022; PI: James Jenkins). HST WFC3 UVIS is also being used to study the eclipse spectrum of LTT 9779 b \citep[GO-16915; PI: Michael Radica; ][]{radica_ltt_uv} and these data may further constrain the eclipse-depth variability as well as helping to determine the albedo. Given the HST observations used here were taken roughly a year apart, the conclusions of these eclipse studies may indicate that, while the G102 and G141 data have similar transit depths in their overlapping spectral region, dynamic changes to the planet may mean the spectra are not compatible.

\subsection{The Orbit of LTT\,9779\,b}\label{sec:orbit}

\begin{figure}
    \centering
    \includegraphics[width=\columnwidth]{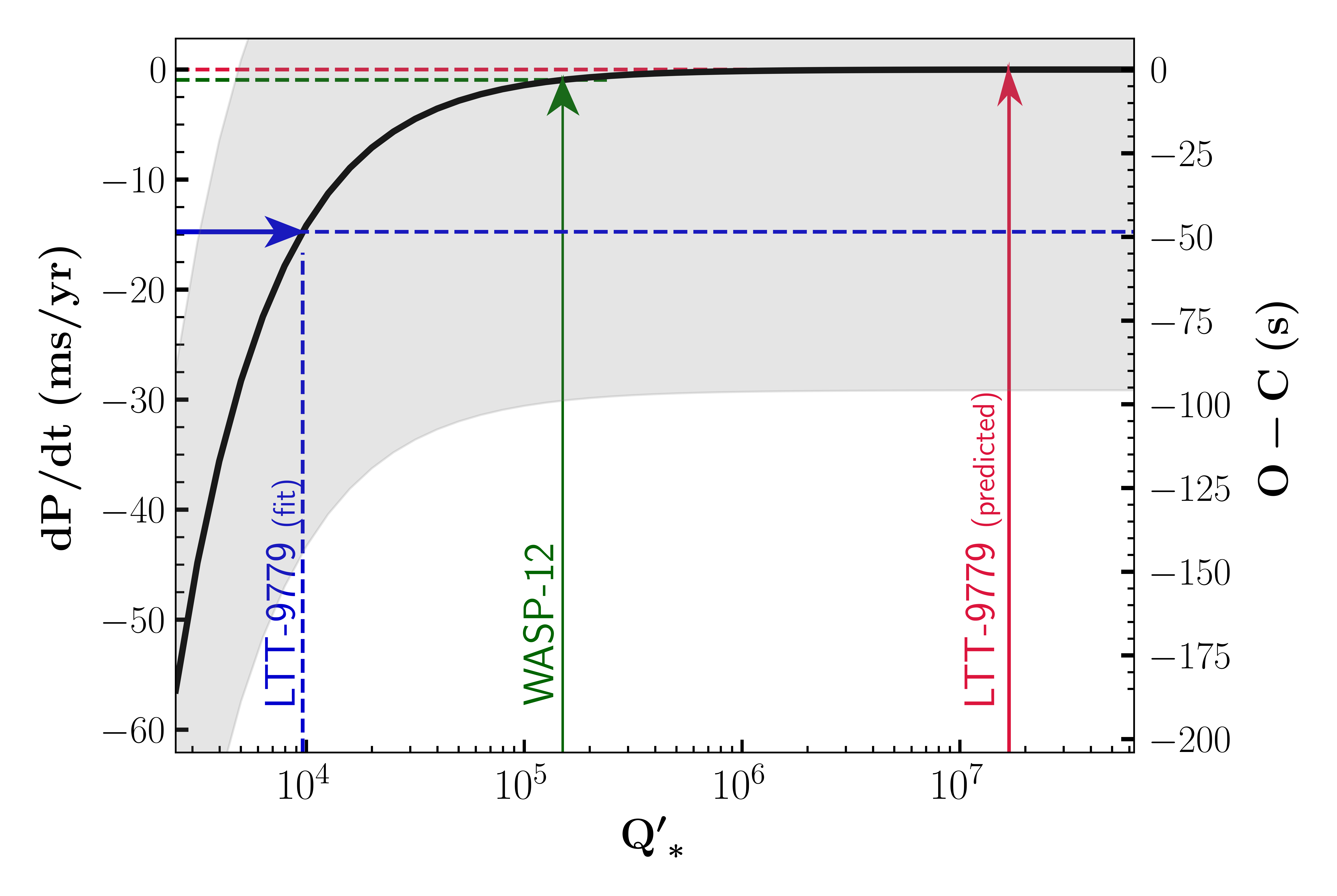}
    \caption{The orbital period derivative in milliseconds per year expected for LTT\,9779\,b given different values of the modified stellar tidal quality factor $Q^{'}_*$. The right-hand vertical axis shows the expected O--C signal over the current observational baseline and the grey shaded region represents the average timing uncertainty in the data. Three values of $Q^{'}_*$ are highlighted: (1) derived with Eq. \ref{equation:dpdt} from the best-fit orbital decay rate when the Spitzer transit timings were included, (2) predicted with Eq. \ref{equation:Q} from the tidal forcing period of the system, and (3) the value for WASP-12 from \citet{wong_w12}.}
    \label{fig:Q}
\end{figure}

While we do not find compelling evidence in the data to suggest that LTT\,9779\,b is undergoing orbital decay, we explore the magnitude of tidal dissipation of energy within the star necessary to yield an observable signal. Assuming the constant-phase lag model for tidal evolution suggested by \citet{goldreich_q_1966}, the expected decay rate is given by:

\begin{eqnarray}
\frac{dP}{dt} &=& -\frac{27\pi}{2Q'_*}\left(\frac{M_p}{M_*}\right)\left(\frac{R_*}{a}\right)^5\rm
\label{equation:dpdt}
\end{eqnarray}
where $M_p$ is the planet's mass, $M_*$ is the star's mass, $R_*$ is the star's radius, $a$ is the semi-major axis of the planet's orbit and $Q^{'}_*$ is the modified tidal quality factor. From our fit of the orbital decay model this yields a $Q^{'}_*$ of $1.5 \times 10^4$ (without Spitzer transits) or $2.1 \times 10^3$ (with Spitzer transits), both far outside the range of $10^5<Q^{'}_*<10^8$ expected given the current theory of energy dissipation within the convective envelopes of cool stars \citep{penev_empirical_2018} and two orders of magnitude smaller than the value of $1.5 \times 10^5$ derived for WASP-12 \citep{patra_apparently_2017, yee_orbit_2019, wong_w12}. Based on this analysis and the strong preference of the Bayes factor for a linear ephemeris, we conclude that the best-fit orbital decay rates from the data do not warrant further consideration.

\citet{penev_empirical_2018} derived an empirical model for $Q^{'}_*$ given the tidal forcing period of the star-planet system $P_{\text{tide}}$ from an analysis of all known exoplanet systems with $M_p>0.1M_{\text{Jup}}$, $P<3.5$ days, and $T_{eff,*}<6100$\,K, the parameter space in which the LTT\,9779 system resides. They found,

\begin{eqnarray}
Q^{'}_* &=& \max\left[{\frac{10^6}{(P_{\text{tide}}/\text{days})^{3.1}}},{10^5}\right]\rm~where\\
P_{\text{tide}} &=& 1/{2\left(\frac{1}{P_{\text{orb}}}-\frac{1}{P_{\text{spin}}}\right)}\rm.
\label{equation:Q}
\end{eqnarray}

Based on this result, we predict $Q^{'}_*$ for LTT\,9779 by adopting a stellar rotation period of 45 days, as found in \citet{jenkins_ltt9779}. Taking this rotation period at face value requires the assumption that the orbit of LTT\,9779\,b is aligned with the stellar rotation axis and so this is, in effect, an upper limit. With this model and assumption, we predict that $Q^{'}_*=1.7 \times 10^7$ for LTT\,9779 which corresponds to an orbital decay rate of $-0.0085$ ms$\text{yr}^{-1}$ for LTT\,9779\,b. Figure \ref{fig:Q} shows that if orbital decay were occurring at this rate, the O--C expected within the current observational baseline would be far below the average uncertainty of the mid-transit times. In the near-future, more data will be available from CHEOPS, HST, JWST and TESS observations. By time the next TESS data for this planet is taken in Sector 69, LTT\,9779\,b will have orbited its host star around 2300 times since the first detection of a transit. If we assume the $Q^{'}_*$ predicted from the tidal forcing period, the O--C would be a mere $-0.05$ seconds by that time. Thus, despite the very short orbital period of LTT\,9779b it is unlikely that, if occurring, tidal orbital decay could be detected.

Likewise, our fit to the timing data suggests that there is only evidence for orbital precession when the Spitzer data is included and that the best-fit apsidal precession model seems to be exploiting the large gaps in the coverage of the transit and eclipse data. Nevertheless, we explore the potential magnitude of precession we might expect for this world, comparing it to preferred value from our fit. We estimate a maximum theoretical apsidal precession rate by summing the contributions of general relativistic effects and tidal and rotational bulges of both the planet and star, using Equations (6), (10), and (12) from \citet{ragozzine_probing_2009}. The magnitude of apsidal precession raised by tidal and rotational bulges depends on the Love number $k_2$ of the perturbing body, thus we assume a value of $k_{2*}=0.03$ for LTT\,9779, typical of sun-like main sequence stars \citep{ragozzine_probing_2009}, and $k_{2p}=0.3$ for LTT\,9779\,b. We chose the latter as a generous value for a hot Neptune with a core mass fraction of $\sim$0.9 \citep{jenkins_ltt9779} given the results of the case study of the Hot Neptune GJ 436\,b by \citet{kramm_degeneracy_2011}, noting that lower values would reduce the predicted rates of precession. Given that we additionally assume the upper limit of orbital eccentricity ($e<0.058$) from \citet{jenkins_ltt9779}, we effectively calculate the maximum possible precession rate from the aforementioned effects. Our calculations yield:


\begin{equation}
    \begin{aligned}
    \dot{\omega}_{\text{tot}} &=& \dot{\omega}_{\text{tide},p} + \dot{\omega}_{\text{tide},*} + \dot{\omega}_{\text{rot},p} + \dot{\omega}_{\text{rot},*} + \dot{\omega}_{\text{GR}} \\
    &=& (4.0\times10^{-5} + 1.5\times10^{-7} + 2.6\times10^{-6}\\
    &+& 3.6\times10^{-8} + 1.1\times10^{-5})\,rad/E \\
    &\approx& 5.4\times10^{-5} rad/E
    \end{aligned}
    \label{eq:prediction}
\end{equation}

where $\dot{\omega}_{\text{tide}}$ and $\dot{\omega}_{\text{rot}}$ refer to the contribution from the tidal and rotational bulges, and $\dot{\omega}_{\text{GR}}$ being from general relativistic effects. The cumulative precession rate of $5.4\times10^{-5}$ rad/E is two orders of magnitude below our best-fit model ($d\omega/dE = 3.1\times10^{-3}$ rad/E). Given that we have assumed characteristics that would maximise the theoretical apisdal precession rate, we conclude that none of the above effects could be the source of the timing variation seen in Figure \ref{fig:oc}.

\begin{figure}
    \centering
    \includegraphics[width=\columnwidth]{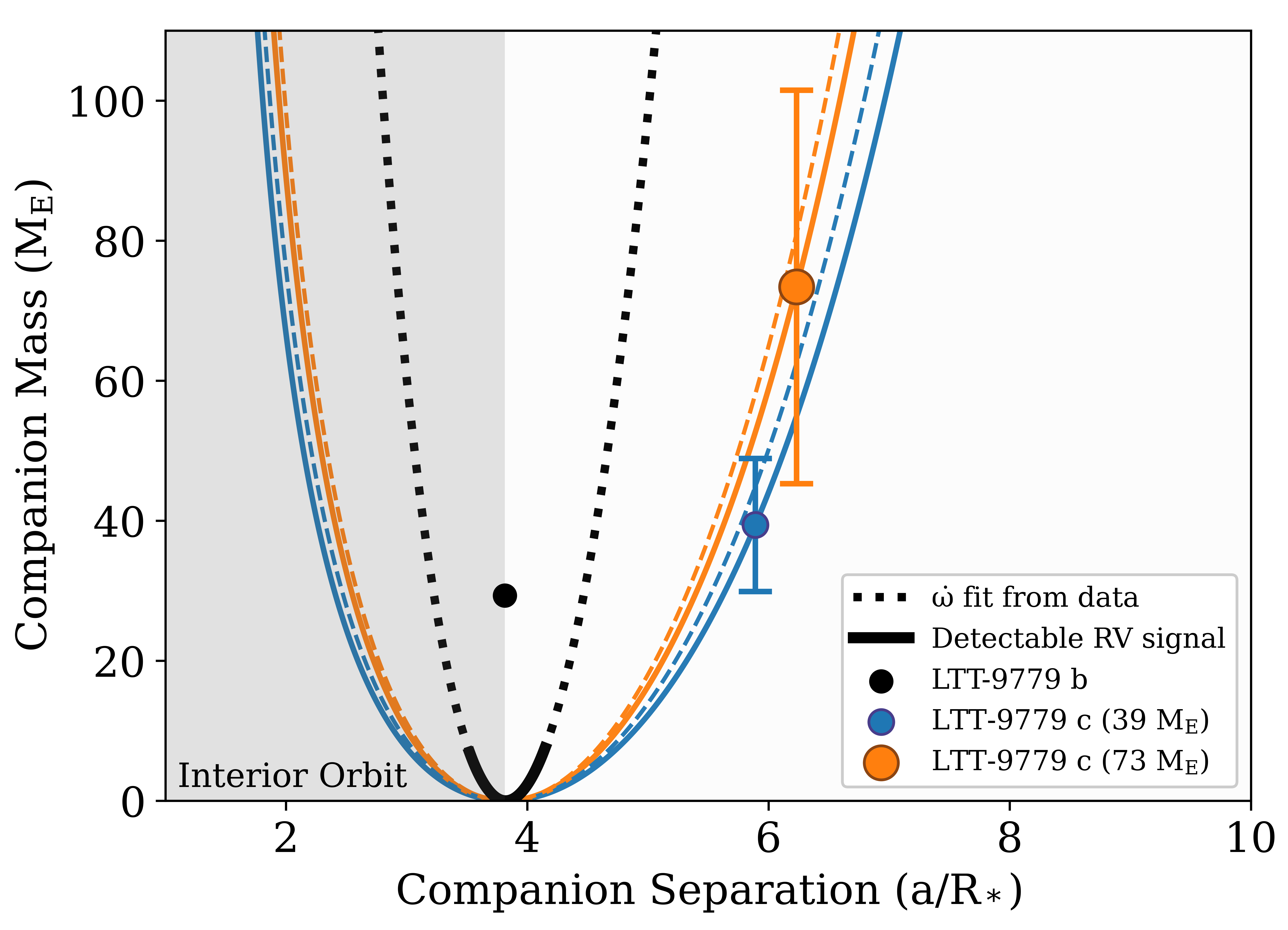}
    \caption{One source of orbital precession could be other planets within the system. We show the possible parameter space for the mass and orbital separation of a companion planet (LTT\,9779 c) for three possible precession rates: the best-fit value from our model fit (black) and those of the two candidate solutions identified in \citet{pearson_cp} (blue and orange). The blue and orange dashed lines are the same as the latter but corrected for the expected precession from other effects calculated in Equation \ref{eq:prediction}. The solid part of the black line represents the possible mass-separation solutions for a companion planet that would provide the best-fit apsidal precession rate from the timing data while remaining undetected in the radial velocity data (K$<$5 m/s) given the residuals of our fit (see Figure \ref{fig:oc}).}
    \label{fig:comp_precession}
\end{figure}

\begin{figure}
    \centering
    \includegraphics[width=\columnwidth]{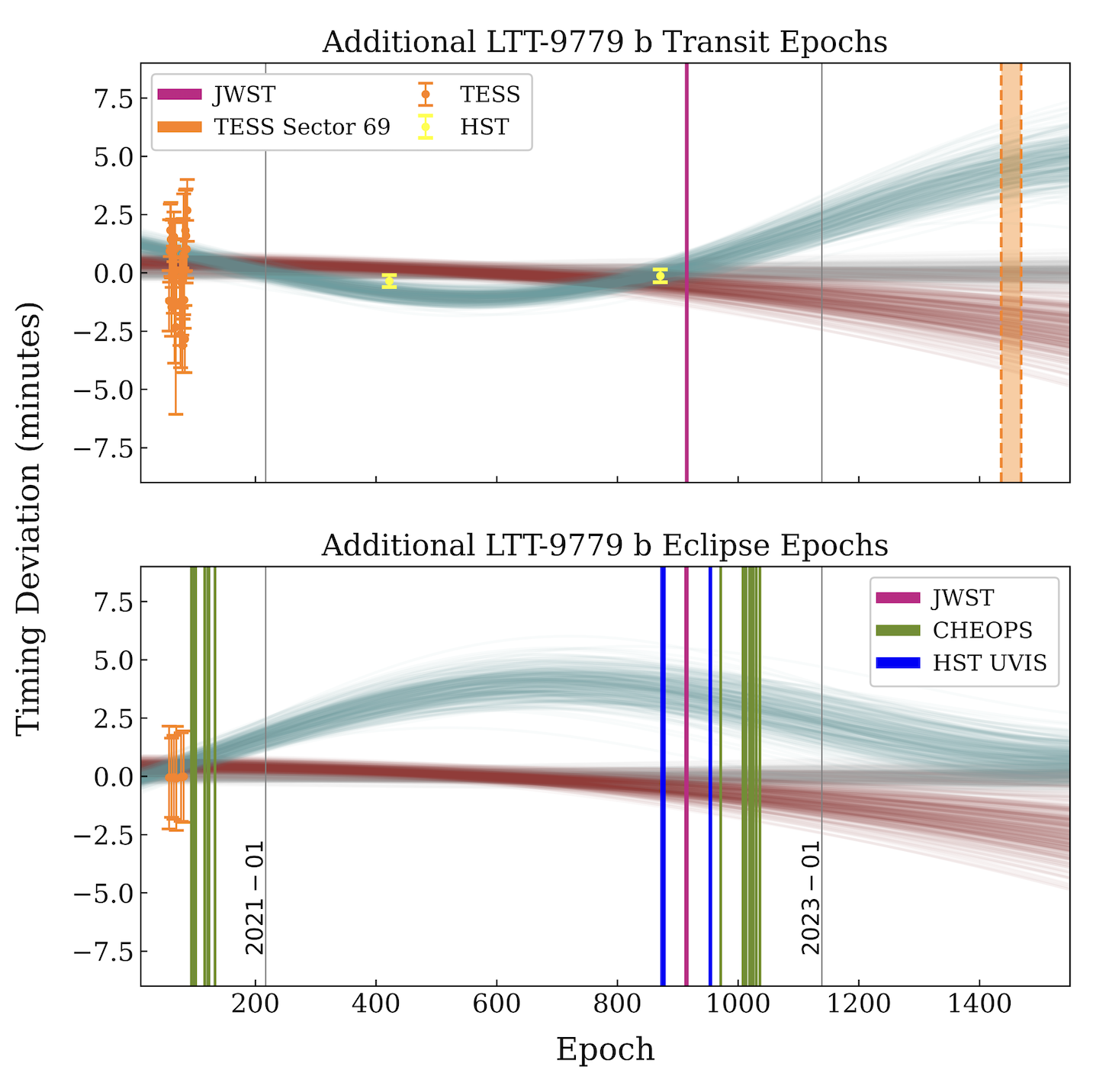}
    \caption{Epochs covered by existing data with CHEOPS, JWST and HST WFC3 UVIS, as well as those that will be covered in TESS Sector 69. Over-plotted are our best-fit models when the Spitzer transit timings are included in the analysis. With these data, the apsidal precession model is likely to be conclusively ruled out while the decay model will still be harder to distinguish from the linear period, although stronger constraints on the maximum decay rate could be placed.}
    \label{fig:oc_zoom}
\end{figure}

We also consider the possibility that apsidal precession may be induced by an additional body within the system. Before the confirmation of LTT\,9779\,b by \citet{jenkins_ltt9779}, there were suggestions of other planets within the LTT 9779 system \citep{pearson_cp}, the best solutions being a 39.4$\pm$9.5 M$_\oplus$ planet on a 1.516 day orbit and a 73.4$\pm$28.1 M$_\oplus$ planet on a 1.65 day orbit. Using Equation (8) from \citet{heyl_using_2007}, we calculate the expected apsidal precession rate raised by these proposed companions. However, we again find that the predicted precession rate(s) would be far below the best-fit value from our timing models (see Figure \ref{fig:comp_precession}), and note that any companion inducing such a strong precession rate would be easily detectable in the radial velocity data. Hence, we find no reasonable way to explain the precession rate fit when including the Spitzer transits, concluding that it is not a viable orbital solution and that it is likely due to poor sampling of the transit and eclipse epochs since the discovery of LTT\,9779\,b. The combination of current and future transit and eclipse data from CHEOPS, HST WFC3 UVIS, JWST, and TESS should be capable of providing definitive evidence against the precession model, as highlighted in Figure \ref{fig:oc_zoom}.

\section{Conclusions}

We use open-source codes to reduce and analyse the HST WFC3 G102 \& G141 transmission spectrum of the ultra-hot Neptune LTT\,9779\,b. However, the constraints placed on the atmospheric parameters from these data are poor, with the best-fit value of many parameters sometimes simply being the centre of the prior. Nevertheless, our retrievals on these data prefer the presence of H$_2$O, CO$_2$ and FeH and the significance of these detections varies from 2-3$\sigma$. On the other hand, the best-fit values for the molecular abundances are higher than expected from chemical equilibrium models. While we find no evidence for excess absorption around the Helium triplet, our modelling shows this is expected given the stellar irradiation. We find little evidence to suggest that LTT\,9779\,b's orbit is decaying, but this is unsurprising given the expected magnitude of the orbital decay and its small effect on the transit times. While a model for apsidal precession can preferably fit the timing data, it yields a solution which cannot be physically explained and the detection hinges purely on two transit timings from Spitzer. Therefore, while these HST data shed further light on the nature of LTT\,9779\,b, much remains to be discovered and many questions are left unanswered. Analysis of data from CHEOPS, HST WFC3 UVIS, and JWST NIRISS should begin to resolve some of these outstanding queries to help us understand this rare and intriguing world.

\section*{Acknowledgements}

BE is grateful to Wilson Joy Skipper for coordinating the HST programme GO-16457, under which these HST WFC3 observations were acquired. QC is funded by the European Space Agency under the 2022 ESA Research Fellowship Program. We acknowledge support from ESA through the Science Faculty - Funding reference ESA-SCI-SC-LE-117. NS acknowledges support from the PSL Iris-OCAV project, and from NASA (Grant \#80NSSC19K0336). This project has also received funding from the European Research Council (ERC) under the European Union's Horizon 2020 research and innovation programme (grant agreements No 758892, ExoAI, and No 817540, ASTROFLOW) and from the Science and Technology Funding Council (STFC) grant ST/S002634/1 and ST/T001836/1 and from the UK Space Agency grant ST/W00254X/1.\\

\textbf{Software:} Iraclis \citep{tsiaras_hd209}, TauREx3 \citep{al-refaie_taurex3}, TauREx GGChem \citep{woitke_ggchem,taurex3_chem}, ASteRA \citep{thompson_ret_spots}, ExoTETHyS \citep{morello_exotethys}, PyLightcurve \citep{tsiaras_plc}, RadVel \citep{fulton_radvel}, Multinest \citep{Feroz_multinest,buchner_multinest}, Astropy \citep{astropy}, h5py \citep{hdf5_collette}, emcee \citep{emcee}, Matplotlib \citep{Hunter_matplotlib},  Pandas \citep{mckinney_pandas}, Numpy \citep{oliphant_numpy}, SciPy \citep{scipy}, corner \citep{corner}.\\

\textbf{Data:} This work is based upon publicly-available observations taken with the NASA/ESA Hubble Space Telescope as part of proposals GO-16457 \citep[PI: Billy Edwards; ][]{edwards_prop} and GO-16166 \citep[PI: Kevin France; ][]{france_prop}. These observations were facilitated by the Space Telescope Science Institute, which is operated by the Association of Universities for Research in Astronomy, Inc., under NASA contract NAS 5–26555. The data were obtained from the Mikulski Archive for Space Telescopes (MAST) at the Space Telescope Science Institute. The specific observations analyzed can be accessed via \dataset[DOI: 10.17909/dr7t-g194]{https://doi.org/10.17909/dr7t-g194}. We are thankful to those who operate the Hubble Space Telescope and the corresponding archive, the public nature of which increases scientific productivity and accessibility \citep{peek2019}. 

\section*{Appendix}

\begin{figure*}
    \centering
    \includegraphics[width=\textwidth]{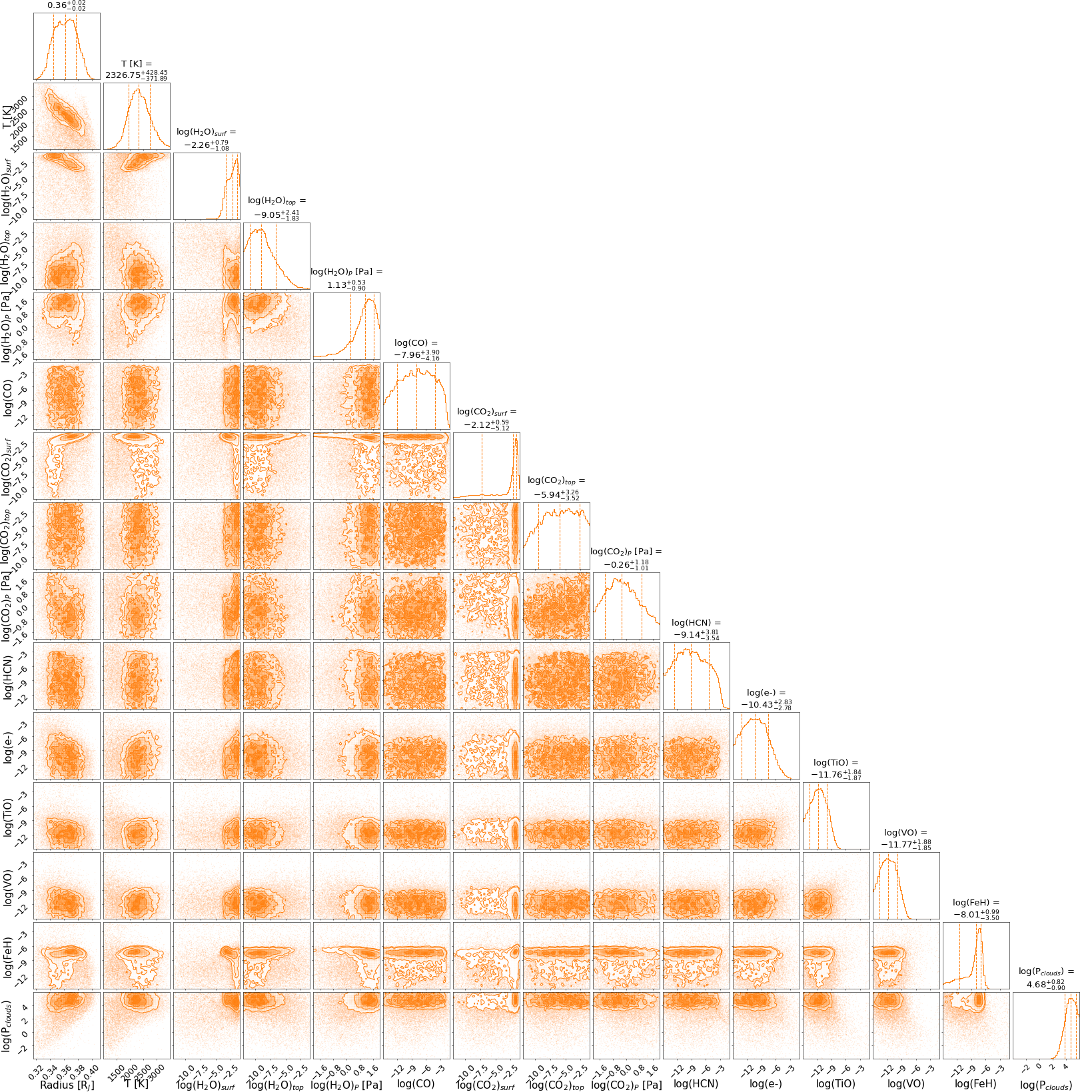}
    \caption{Posterior distributions for our preferred free chemistry retrieval which included an isothermal temperature profile and two-layer abundances for H$_2$O and CO$_2$. We note that, for many parameters, the best-fit value is simply the mean of the prior.}
    \label{fig:free_chem_post}
\end{figure*}

\begin{figure*}
    \centering
    \includegraphics[width=\textwidth]{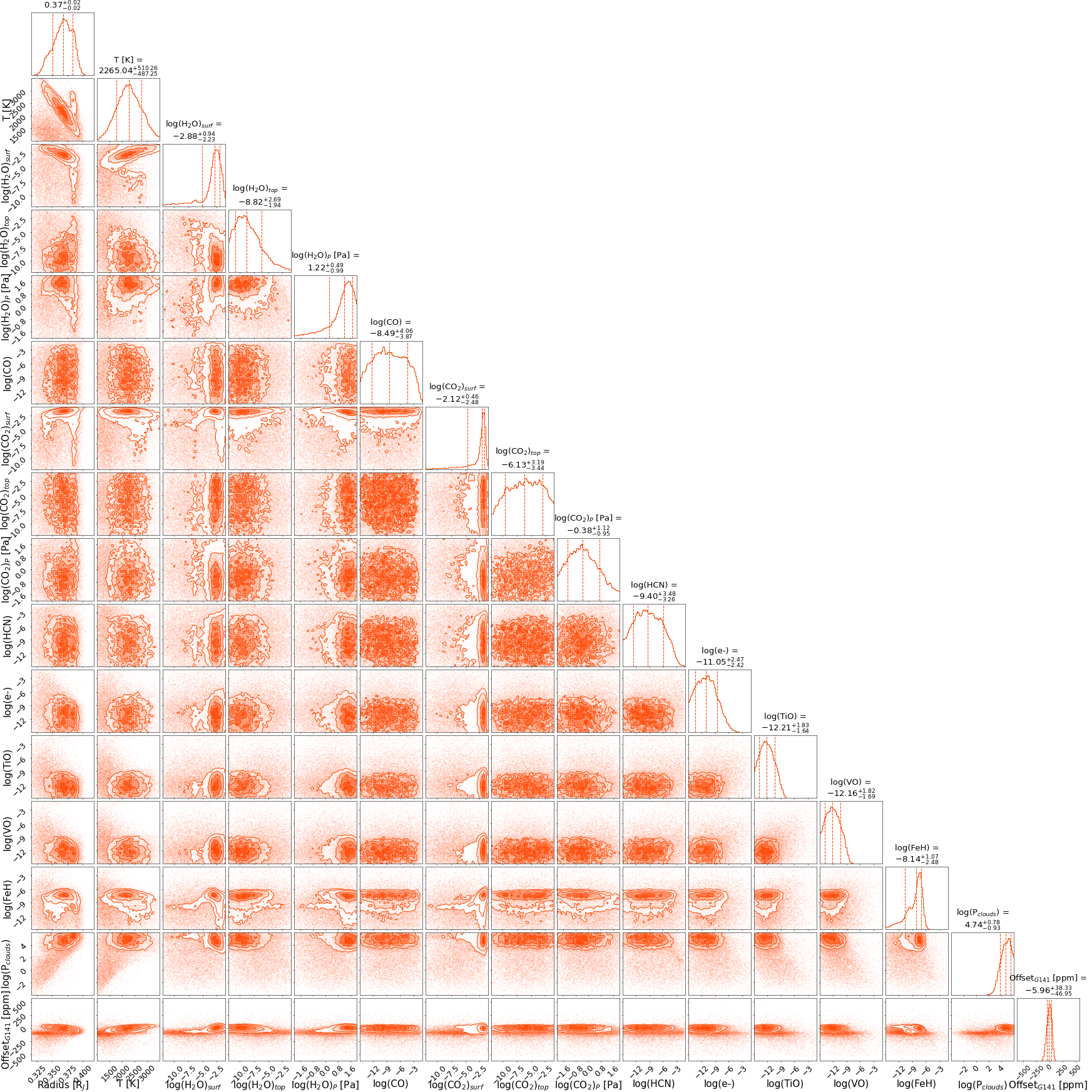}
    \caption{Posterior distributions for the free chemistry retrieval which included an offset in the HST WFC3 G141 data. The retrieval converges to a solution which has a slight offset, but the "no offset" solution is well within the 1 $\sigma$ uncertainties. Furthermore, the Bayesian evidence for this retrieval is significantly less than that of the same retrieval minus an offset parameter. The difference in the Bayesian evidence suggests an offset between the datasets is rejected at the 3.7 $\sigma$ level.}
    \label{fig:free_chem_off_post}
\end{figure*}

\begin{figure*}
    \centering
    \includegraphics[width=\textwidth]{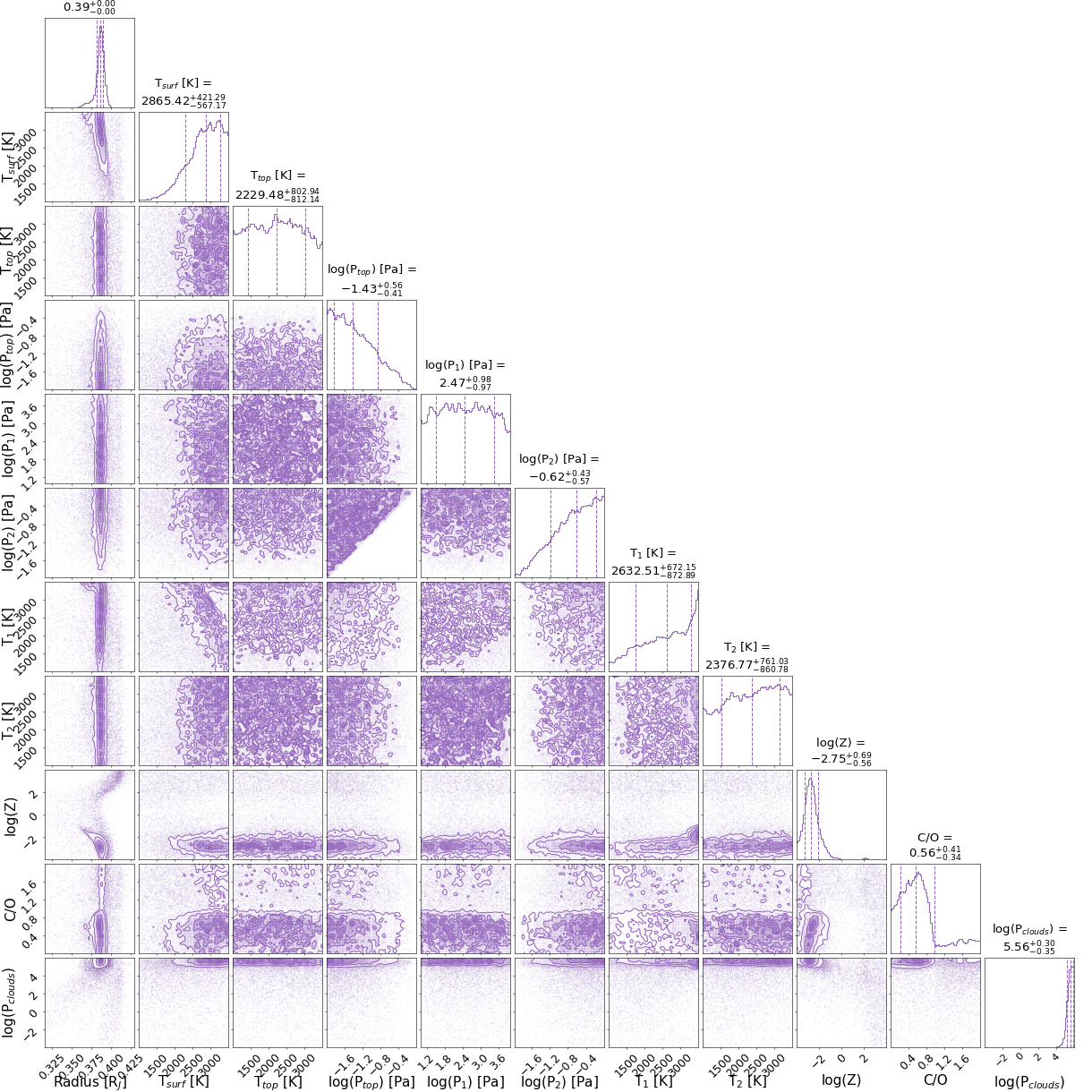}
    \caption{Posterior distributions for our preferred chemical equilibrium retrieval which included a non-isothermal temperature profile. We note that this retrieval was not preferred over the free chemistry model given in Figure \ref{fig:free_chem_post}, with the Bayesian evidence preferring the latter to 2.4 $\sigma$. Additionally, while the non-isothermal temperature profile led to a preferable fit, it did not led to a well-constrained TP profile.}
    \label{fig:chem_eq_post}
\end{figure*}

\begin{figure*}
    \centering
    \includegraphics[width=\textwidth]{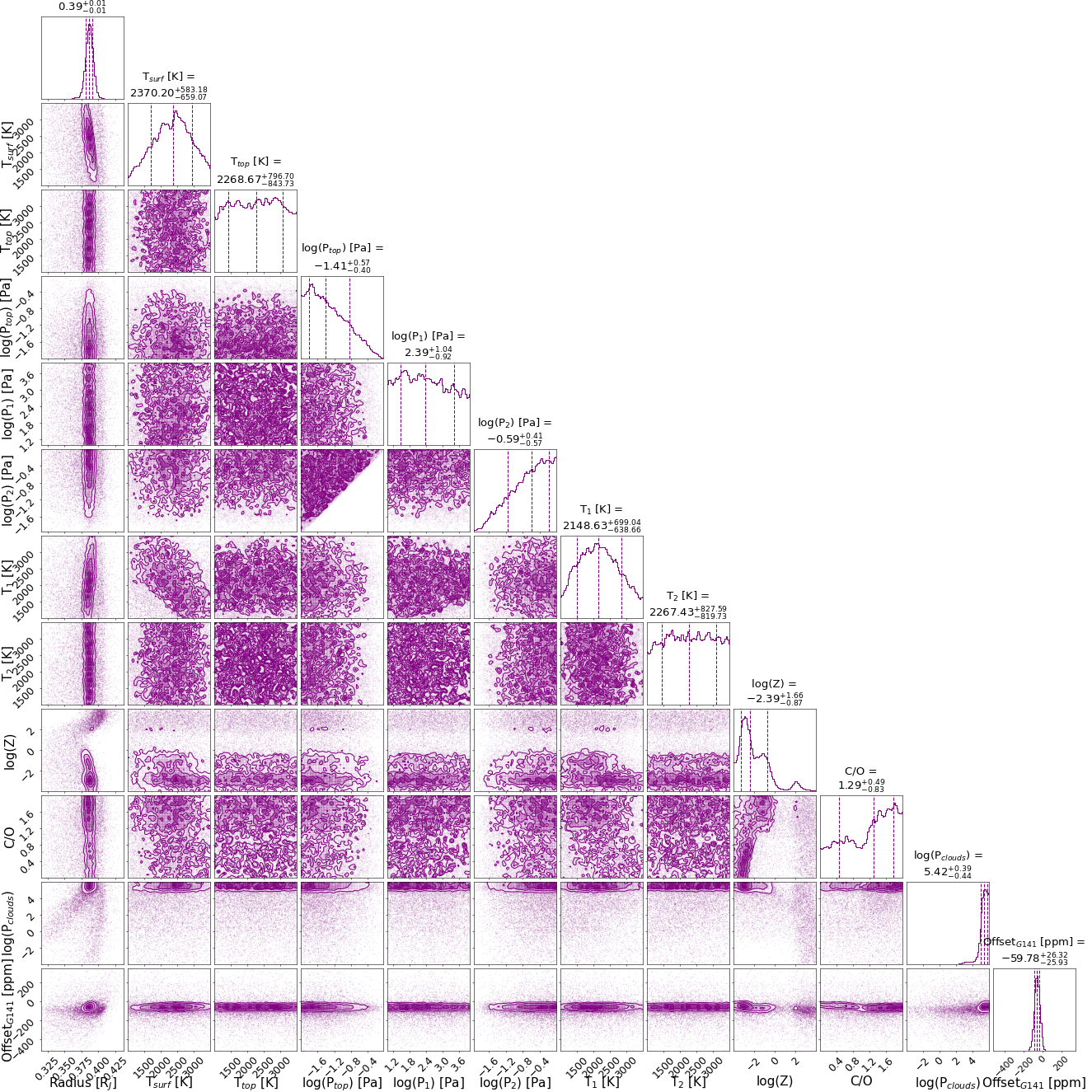}
    \caption{Posterior distributions for the chemical equilibrium retrieval which included an offset in the HST WFC3 G141 data. The retrieval converges to a solution which has a significant offset. However, the Bayesian evidence for this retrieval is significantly less than that of the same retrieval minus an offset parameter. The difference in the Bayesian evidence suggests an offset between the datasets is rejected at the 3.4 $\sigma$ level, the same degree to which the free chemistry retrievals prefer a lack of offset.}
    \label{fig:chem_eq_off_post}
\end{figure*}

\begin{figure*}
    \centering
    \includegraphics[width=\textwidth]{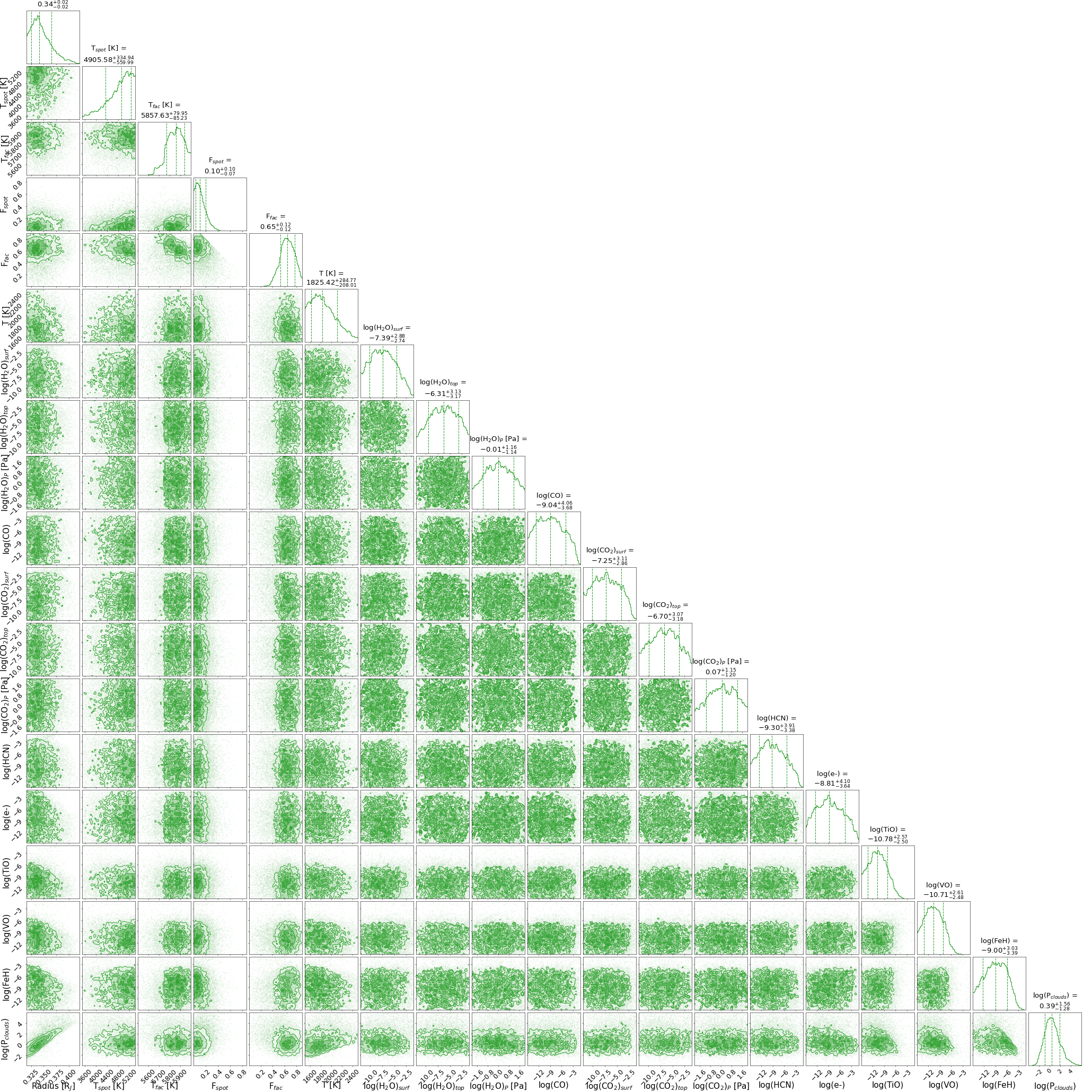}
    \caption{Posterior distributions for our free chemistry retrieval which accounted for spots and faculae. The retrieval prefers a solution without an atmospheric detection for LTT\,9779\,b and instead favours a very large, and unlikely, faculae covering fraction.}
    \label{fig:spotty_post}
\end{figure*}

Here, we provide the posteriors for several of the retrievals conducted in this study. Figure \ref{fig:free_chem_post} shows the posteriors for our preferred free chemistry retrieval while the free chemistry retrieval with an offset between the HST WFC3 G102 and G141 datasets is shown in Figure \ref{fig:free_chem_off_post}. Likewise, Figures \ref{fig:chem_eq_post} and \ref{fig:chem_eq_off_post} show the posteriors from the chemical equilibrium retrievals with and without an offset, respectively. Finally, in Figure \ref{fig:spotty_post} we shown the posteriors from our retrieval which included the effects of spots and faculae.

Table \ref{tab:para} gives the stellar and planetary parameters used in this study while the HST WFC3 G102 and G141 spectra are given in Table \ref{tab:hst_spec}. We also provide the transit and eclipse mid-times used in this study in Tables \ref{tab:tr_mt} and \ref{tab:ec_mt}, respectively. We also provide the best-fit parameters from the linear, decay, and precession models in Table \ref{tab:lin_dec_pre_mod}.

\clearpage

\begin{table}[]
    \centering
    \caption{Stellar and planetary parameters used or derived in this study.}
    \begin{tabular}{cccc} \hline \hline
       Parameter & Unit & Value & Source \\ \hline \hline
       R$_*$ & R$_\odot$ & 0.949 $\pm$ 0.006 & J20 \\
       M$_*$ & M$_\odot$ & 1.02$^{+0.02}_{-0.03}$ & J20 \\
       T$_*$ & K & 5480$\pm$42 & J20 \\
       log(g) & log$_{10}$(cms$^{-2}$) & 4.47$\pm$0.11 & J20 \\
       Fe/H & dex & 0.25$\pm$0.04 & J20 \\ 
       Age & Gyr & 2.0 $^{+1.3}_{-0.9}$\\
       P$_\text{rot}$ & days & $<45$ & J20 \\ \hline
       R$_p$ & R$_\oplus$ & 4.72$\pm$0.23 & J20 \\
       M$_p$ & M$_\oplus$ & 29.32$^{+0.78}_{-0.81}$ & J20 \\
       a/R$_*$ & & 3.877$^{+0.090}_{-0.091}$& J20 \\
       i & deg & 76.39$\pm$0.43 & J20 \\
       e &  & $<0.058$ & J20 \\
       T$_{eq}$ & K & 1978$\pm$19 & J20 \\
       T$_0$ & BJD$_{\rm TDB}$ & 2459043.310602$\pm$0.000090 & TW \\
       P & days & 0.79206410$\pm$0.00000014 & TW \\ \hline 
       \multicolumn{4}{c}{J20: \citet{jenkins_ltt9779}; TW: This Work} \\ \hline \hline
    \end{tabular}
    \label{tab:para}
\end{table}

\begin{table}[]
    \centering
    \caption{The HST WFC3 G102 spectrum extracted in this work and the HST WFC3 G141 spectrum from \citet{transmission_pop}.}
    \begin{tabular}{cccc} \hline \hline
    Wavelength & Transit Depth & Uncertainty & Bandwidth \\ \relax
    [$\mu$m] & [\% ]& [\%] & [$\mu$m] \\ \hline
    0.83750 & 0.1810 & 0.0077 & 0.0250 \\
    0.86250 & 0.1772 & 0.0087 & 0.0250 \\
    0.88750 & 0.1986 & 0.0071 & 0.0250 \\
    0.91250 & 0.1921 & 0.0066 & 0.0250 \\
    0.93750 & 0.1938 & 0.0060 & 0.0250 \\
    0.96250 & 0.1975 & 0.0060 & 0.0250 \\
    0.98750 & 0.2099 & 0.0054 & 0.0250 \\
    1.01250 & 0.2081 & 0.0051 & 0.0250 \\
    1.03750 & 0.2024 & 0.0051 & 0.0250 \\
    1.06250 & 0.1951 & 0.0041 & 0.0250 \\
    1.08750 & 0.1919 & 0.0056 & 0.0250 \\
    1.11250 & 0.2051 & 0.0066 & 0.0250 \\
    1.13750 & 0.1997 & 0.0051 & 0.0250 \\ \hline 
    1.12625 & 0.2175 & 0.0064 & 0.0219 \\
    1.14775 & 0.1970 & 0.0060 & 0.0211 \\
    1.16860 & 0.1963 & 0.0087 & 0.0206 \\
    1.18880 & 0.1917 & 0.0071 & 0.0198 \\
    1.20835 & 0.2002 & 0.0062 & 0.0193 \\
    1.22750 & 0.1956 & 0.0065 & 0.0190 \\
    1.24645 & 0.1953 & 0.0071 & 0.0189 \\
    1.26550 & 0.2163 & 0.0056 & 0.0192 \\
    1.28475 & 0.1921 & 0.0062 & 0.0193 \\
    1.30380 & 0.2013 & 0.0060 & 0.0188 \\
    1.32260 & 0.2036 & 0.0073 & 0.0188 \\
    1.34145 & 0.2054 & 0.0076 & 0.0189 \\
    1.36050 & 0.2090 & 0.0071 & 0.0192 \\
    1.38005 & 0.2136 & 0.0073 & 0.0199 \\
    1.40000 & 0.2093 & 0.0065 & 0.0200 \\
    1.42015 & 0.2006 & 0.0062 & 0.0203 \\
    1.44060 & 0.2112 & 0.0063 & 0.0206 \\
    1.46150 & 0.2097 & 0.0058 & 0.0212 \\
    1.48310 & 0.2103 & 0.0052 & 0.0220 \\
    1.50530 & 0.2180 & 0.0070 & 0.0224 \\
    1.52800 & 0.2149 & 0.0051 & 0.0230 \\
    1.55155 & 0.2074 & 0.0060 & 0.0241 \\
    1.57625 & 0.2116 & 0.0063 & 0.0253 \\
    1.60210 & 0.2143 & 0.0060 & 0.0264 \\
    1.62945 & 0.2203 & 0.0061 & 0.0283 \\ \hline \hline
    \end{tabular}
    \label{tab:hst_spec}
\end{table}

\begin{longtable}{cccc} 
\caption{Transit mid-times utilised in this work.} \label{tab:tr_mt} \\ \hline \hline
    Transit Mid-Time  & Uncertainty & Epoch & Source \\ \relax 
    [BJD$_{\rm TDB}$] & [days] & & \\ \hline \hline
    2458354.21580 & 0.00120 & -870 & TESS (K22) \\
    2458355.00840 & 0.00160 & -869 & TESS (K22) \\
    2458355.79800 & 0.00110 & -868 & TESS (K22) \\
    2458357.38318 & 0.00099 & -866 & TESS (K22) \\
    2458358.17580 & 0.00110 & -865 & TESS (K22) \\
    2458358.96760 & 0.00200 & -864 & TESS (K22) \\
    2458359.75890 & 0.00110 & -863 & TESS (K22) \\
    2458360.55100 & 0.00130 & -862 & TESS (K22) \\
    2458361.34230 & 0.00140 & -861 & TESS (K22) \\
    2458362.13651 & 0.00100 & -860 & TESS (K22) \\
    2458362.92650 & 0.00130 & -859 & TESS (K22) \\
    2458363.71870 & 0.00100 & -858 & TESS (K22) \\
    2458364.51130 & 0.00120 & -857 & TESS (K22) \\
    2458365.30484 & 0.00077 & -856 & TESS (K22) \\
    2458366.09510 & 0.00100 & -855 & TESS (K22) \\
    2458366.88765 & 0.00100 & -854 & TESS (K22) \\
    2458369.26457 & 0.00074 & -851 & TESS (K22) \\
    2458370.05550 & 0.00100 & -850 & TESS (K22) \\
    2458370.84950 & 0.00180 & -849 & TESS (K22) \\
    2458371.63979 & 0.00074 & -848 & TESS (K22) \\
    2458372.43220 & 0.00210 & -847 & TESS (K22) \\
    2458373.22520 & 0.00170 & -846 & TESS (K22) \\
    2458374.01730 & 0.00130 & -845 & TESS (K22) \\
    2458374.80930 & 0.00130 & -844 & TESS (K22) \\
    2458375.60121 & 0.00092 & -843 & TESS (K22) \\
    2458376.39385 & 0.00076 & -842 & TESS (K22) \\
    2458377.18376 & 0.00077 & -841 & TESS (K22) \\
    2458377.97850 & 0.00170 & -840 & TESS (K22) \\
    2458378.76838 & 0.00080 & -839 & TESS (K22) \\
    2458379.56140 & 0.00150 & -838 & TESS (K22) \\
    2458380.35167 & 0.00099 & -837 & TESS (K22) \\
    2458381.14590 & 0.00140 & -836 & TESS (K22) \\
    2458781.13997 & 0.00032 & -331 & Spitzer (C20) \\
    2458783.51684 & 0.00053 & -328 & Spitzer (C20) \\
    2459088.45768 & 0.00090 & 57 & TESS (K22) \\
    2459089.25123 & 0.00093 & 58 & TESS (K22) \\
    2459090.04391 & 0.00078 & 59 & TESS (K22) \\
    2459090.83570 & 0.00110 & 60 & TESS (K22) \\
    2459091.62575 & 0.00087 & 61 & TESS (K22) \\
    2459092.41940 & 0.00100 & 62 & TESS (K22) \\
    2459093.21130 & 0.00120 & 63 & TESS (K22) \\
    2459094.00276 & 0.00100 & 64 & TESS (K22) \\
    2459094.79610 & 0.00074 & 65 & TESS (K22) \\
    2459095.58610 & 0.00170 & 66 & TESS (K22) \\
    2459096.37986 & 0.00079 & 67 & TESS (K22) \\
    2459097.16960 & 0.00260 & 68 & TESS (K22) \\
    2459097.96335 & 0.00095 & 69 & TESS (K22) \\
    2459102.71480 & 0.00130 & 75 & TESS (K22) \\
    2459103.50590 & 0.00100 & 76 & TESS (K22) \\
    2459104.29951 & 0.00077 & 77 & TESS (K22) \\
    2459105.09100 & 0.00100 & 78 & TESS (K22) \\
    2459105.88175 & 0.00080 & 79 & TESS (K22) \\
    2459106.67610 & 0.00150 & 80 & TESS (K22) \\
    2459107.46860 & 0.00180 & 81 & TESS (K22) \\
    2459108.25930 & 0.00084 & 82 & TESS (K22) \\
    2459109.05020 & 0.00100 & 83 & TESS (K22) \\
    2459109.84550 & 0.00120 & 84 & TESS (K22) \\
    2459110.63740 & 0.00140 & 85 & TESS (K22) \\
    2459111.42907 & 0.00086 & 86 & TESS (K22) \\
    2459112.22229 & 0.00092 & 87 & TESS (K22) \\
    2459377.56157 & 0.00018 & 422 & HST (TW) \\
    2459733.19839 & 0.00019 & 871 & HST (TW) \\ \hline
    \multicolumn{4}{c}{C20: \citet{crossfield_ltt}; K22: \citet{kokori_3}}\\
    \multicolumn{4}{c}{TW: This Work}\\ \hline \hline
\end{longtable}

\begin{longtable}{cccc}
    \caption{Eclipse mid-times utilised in this work.} \label{tab:ec_mt} \\ \hline \hline
    Eclipse Mid-Time  & Uncertainty & Epoch & Source \\ \relax 
    [BJD$_{\rm TDB}$] & [days] & & \\ \hline \hline
    2458354.61092 & 0.00205 & -870 & TESS (TW) \\
    2458358.57127 & 0.00115 & -865 & TESS (TW) \\
    2458360.94750 & 0.00123 & -862 & TESS (TW) \\
    2458364.11575 & 0.00150 & -858 & TESS (TW) \\
    2458368.86808 & 0.00102 & -852 & TESS (TW) \\
    2458372.03640 & 0.00164 & -848 & TESS (TW) \\
    2458375.20465 & 0.00234 & -844 & TESS (TW) \\
    2458378.37286 & 0.00111 & -840 & TESS (TW) \\
    2458541.53740 & 0.00110 & -634 & Spitzer (D20) \\
    2458544.70430 & 0.00200 & -630 & Spitzer (D20) \\
    2458550.24910 & 0.00260 & -623 & Spitzer (D20) \\
    2458555.00240 & 0.00130 & -617 & Spitzer (D20) \\
    2458562.13080 & 0.00200 & -608 & Spitzer (D20) \\
    2458563.71590 & 0.00170 & -606 & Spitzer (D20) \\
    2458569.25990 & 0.00280 & -599 & Spitzer (D20) \\
    2458574.80220 & 0.00290 & -592 & Spitzer (D20) \\
    2458780.73960 & 0.00670 & -332 & Spitzer (D20) \\
    2458781.53340 & 0.00130 & -331 & Spitzer (D20) \\
    2458783.11600 & 0.00350 & -329 & Spitzer (D20) \\
    2458783.90490 & 0.00400 & -328 & Spitzer (D20) \\
    2459088.85451 & 0.00153 & 57 & TESS (TW) \\
    2459092.02276 & 0.00118 & 61 & TESS (TW) \\
    2459095.19103 & 0.00125 & 65 & TESS (TW) \\
    2459098.35925 & 0.00155 & 69 & TESS (TW) \\
    2459104.69580 & 0.00132 & 77 & TESS (TW) \\
    2459107.86407 & 0.00136 & 81 & TESS (TW) \\ \hline
    \multicolumn{4}{c}{D20: \citet{dragomir_ltt}; TW: This Work}\\ \hline \hline
\end{longtable}



\begin{table*}[]
    \centering
    \caption{Best-fit parameters from our linear, decay, and precession model fits to the transit and eclipse mid-times and radial velocity data of LTT\,9779\,b. Due to the significant O-C seen in the Spitzer transit mid-times, but not the Spitzer eclipse mid-times, we show the results for fits with and without these transit mid-times, preferring the latter as discussed in Section \ref{sec:updated_ephemeris}.}
    \begin{tabular}{ccccc} \hline \hline
     Model & Parameter & Unit & Without Spitzer Transits & With Spitzer Transits \\ \hline \hline
     \multirow{2}{*}{Linear} & P & [days] & 0.79206406 $\pm$ $0.00000014^\dagger$ & 0.79206380 $\pm$ 0.00000014\\
     & $t_0$ & [BJD$_{\rm TDB}$] & 2459043.310621 $\pm$ $0.000088^\dagger$ & 2459043.310895 $\pm$ 0.000086\\ \hline

     \multirow{4}{*}{Decay} & P & [days] & 0.79206404 $\pm$ 0.00000014 & 0.79206376 $\pm$ 0.00000014\\
      & $t_0$ & [BJD$_{\rm TDB}$] & 2459043.31067 $\pm$ 0.00015 & 2459043.31122 $\pm$ 0.00014\\
      & dP/dE & [days epoch$^{-1}$] & $-2.4\times10^{-10}$ $\pm$ $5.6\times10^{-10}$ & $-1.71\times10^{-9}$ $\pm$ $0.54\times10^{-9}$\\
      & dP/dt & [ms yr$^{-1}$] & $-9$ $\pm$ 22 & $-68$ $\pm$ 21 \\ \hline

     \multirow{4}{*}{Precession} & P & [days] & 0.79206409 $\pm$ 0.00000015 & 0.79206476 $\pm$ 0.00000023\\
      & $t_0$ & [BJD$_{\rm TDB}$] & 2459043.310601 $\pm$ 0.000093 & 2459043.311163 $\pm$ 0.000092\\
      & $\omega_0$ & [rad] & 3.8 $\pm$ 2.1 & 4.358 $\pm$ 0.070\\
      & d$\omega$/dE & [rad epoch$^{-1}$] & 0.0030 $\pm$ 0.0023 & 0.00313 $\pm$ 0.00015 \\
      & e & & 0.00018 $\pm$ 0.00042 & 0.00623 $\pm$ 0.00082\\ \hline
     \multicolumn{5}{c}{$^\dagger$Preferred ephemeris solution}\\
     \hline \hline
    \end{tabular}
    \label{tab:lin_dec_pre_mod}
\end{table*}

\bibliography{main}{}
\bibliographystyle{aasjournal}

\end{document}